\documentclass[a4paper,11pt]{article}
\usepackage{jcappub} 
\usepackage{lineno}
\usepackage{fontawesome}
\usepackage{comment}

\title{Forecast cosmological constraints from the number counts of Gravitational Waves events}

\author[a]{Giovanni Antinozzi,}
\author[b,c]{Matteo Martinelli,}
\author[b,d]{and Roberto Maoli}

\affiliation[a]{SISSA, via Bonomea 265, 34136 Trieste, Italy}
\affiliation[b]{INAF - Osservatorio Astronomico di Roma, via Frascati 33, 00040 Monteporzio Catone (Roma), Italy}
\affiliation[c]{INFN - Sezione di Roma, Piazzale Aldo Moro, 2 - c/o Dipartimento di Fisica, Edificio G. Marconi, I-00185 Roma, Italy}
\affiliation[d]{Dipartimento di Fisica, Sapienza Universit\`a di Roma, Piazzale Aldo Moro 2, I-00185 Roma, Italy}

\emailAdd{giovanni.antinozzi@sissa.it}
\emailAdd{matteo.martinelli@inaf.it}
\emailAdd{roberto.maoli@roma1.infn.it}

\abstract{We present a forecast for the upcoming Einstein Telescope (ET) interferometer with two new methods to infer cosmological parameters. We consider the emission of Gravitational Waves (GWs) from compact binary coalescences, whose electromagnetic counterpart is missing, namely Dark Sirens events. 
Most of the methods used to infer cosmological information from GW observations rely on the availability of a redshift measurement, usually obtained with the help of external data, such as galaxy catalogues used to identify the most likely galaxy to host the emission of the observed GWs.
Instead, our approach is based only on the GW survey itself and exploits the information on the distance of the GW rather than on its redshift. Since a large dataset spanning the whole distance interval is expected to fully represent the distribution, we applied our methods to the expected ET's far-reaching measuring capabilities. We simulate a dataset of observations with ET using the package \texttt{darksirens}, assuming an underlying $\Lambda$CDM cosmology, and including the possibility to choose between three possible Star Formation Rate density (SFR) models, also accounting for possible population III stars (PopIII). We test two independent statistical methods: one based on a likelihood approach on the theoretical expectation of observed events, and another applying the \emph{cut-and-count method}, a simpler method to compare the observed number of events with the predicted counts. Both methods are consistent in their final results, and also show the potential to distinguish an incorrect SFR model from the data, but not the presence of a possible PopIII. Concerning the cosmological parameters, we find instead that ET observations by themselves would suffer from strong degeneracies, but have the potential to significantly contribute to parameter estimation if used in synergy with other surveys.}

\begin{document}
\maketitle
\flushbottom

\section{Introduction}
\label{sec:intro}

Gravitational Waves Astronomy has recently emerged as one of the most fruitful predictions of General Relativity, opening new observational windows onto the Universe. Since the rigorous theoretical formulation in 1916, the first direct measurement was achieved only after a century of experimental effort, marking September $14^{th}$ 2015 as the birth of a new field of research in Physics \cite{LIGOScientific:2016aoc}.

Since this first detection, numerous other events have been observed, thanks to the LIGO, Virgo and KAGRA collaborations that provided the latest catalogue of observed GW events \cite{LIGOScientific:2021djp}. The list of observed sources of GWs was however restricted to coalescences of combinations of Black Holes (BHs) and Neutron Stars (NSs), until very recently. On June $28^{th}$ 2023 the NANOGrav international collaboration announced the evidence for a Gravitational Wave Background (GWB), the long sought after mixed collection of low frequency GW source emissions \cite{NANOGrav:2023gor}. Even though the dominant contribution to the GWB has to probably be attributed to a population of supermassive black hole binaries, other cosmologically relevant sources may be hidden in the same GW emission frequency band, such as primordial GWs due to inflation, scalar-induced GWs and GWs from processes resulting from cosmological phase transitions \cite{NANOGrav:2023gor}.

Such new probes have recently become of paramount importance in cosmology, since increased precision brought to light the existence of tensions between measurement of cosmological parameters obtained from different observations (see e.g. \cite{Perivolaropoulos:2021jda,Abdalla:2022yfr}). The direct measurement of gravitational waves is exactly the new probe that cosmology needed: an abundance of previously concealed information, which could be used in conjunction with electromagnetic emissions to infer distance measurements on cosmological scales.
However, observations of GWs do not provide a redshift value $z$, since what we observe are not electromagnetic waves, and this has led to the development of a multitude of methods to obtain such information. 

In the rare case where we can measure the redshift of the binary through the GWs electromagnetic counterpart, we call these events \emph{Standard Sirens} or \emph{Bright Sirens} as opposed to \emph{Dark Sirens} where $z$ has to be found with alternative techniques. 

Furthermore, Dark Sirens can be classified as a function of the method used to find z: \emph{Statistical Sirens}, where the sky patch localized by GW events is scanned for galaxies or galaxy clusters whose redshifts are combined to obtain the most likely value of the true redshift \cite{Schutz:1986gp,DelPozzo:2011vcw,Chen:2017rfc,Yu:2023ico}; \emph{Spectral Sirens}, where the independence of the mass spectrum distribution from redshift can be exploited to break the mass-redshift degeneracy \cite{Taylor:2012db}; \emph{Love Sirens}, where source frame masses of a neutron star binary system are obtained from the direct measurement of its tidal deformability, hence breaking again the mass-redshift degeneracy \cite{Yagi:2013sva,DelPozzo:2015bna,Wang:2020xwn,Chatterjee:2021xrm,Jin:2022qnj,Ghosh:2022muc}; \emph{Gray Sirens}, when a BH-NS system is doubly used as a Dark Siren and as a Bright Siren \cite{Gupta:2022fwd}; and finally we mention a term often used: \emph{Golden Sirens}, i.e. well localized single event Dark Sirens such that the resulting Hubble constant estimate can resolve the Hubble tension (i.e. with sub-percent precision) \cite{Borhanian:2020vyr}.

In this work we focus instead on the possibility of obtaining information from GW surveys without the need to find the events' redshift. This necessarily requires one to work in a statistical framework and therefore it requires extremely high numbers of events to obtain cosmological information, but such a disadvantage is compensated by the fact that one can exploit all observations obtained by GW surveys, without being limited to a subset where the redshift information can be obtained. 

One can obtain theoretical predictions for the number of observed events and their distribution in luminosity distance, a quantity that can be obtained from observations, which can allow one to extract cosmological information from GW surveys without the need to depend only on rare events or external data \cite{Ding:2018zrk,Leandro:2021qlc}.

In this paper, we rely on the counts of GW events as our main source of information, building upon previous works where this observable was used to distinguish two different BH populations, namely astrophysical and primordial \cite{Martinelli:2022elq}. Similar approaches have been taken before \cite{Ding:2018zrk,Leandro:2021qlc}, but we account differently for cosmological degeneracies, without including priors on the matter content of the Universe $\Omega_{m}$, and we also apply the \emph{cut-and-count method} \cite{Martinelli:2022elq}, initially introduced for primordial black holes, but applied here for cosmological inference. Furthermore, we also explore the impact of astrophysical assumptions on the results obtained with these methods.

In order to pursue this approach, it is crucial for a high number of events to be available, as the method requires observations to trace the overall distribution of events. For such a reason, we decided not to work with currently available data, but rather to investigate next generation surveys, which will provide a significant increase in event statistics, focusing in particular on the planned European Space Agency Einstein Telescope (ET) \cite{Maggiore:2019uih}, for which the projected number of detectable events will satisfy our dataset size requirements.

The paper is organized as follows. In \autoref{sec:gw_obs} we briefly review the main equations connecting the observable quantities of GW events to cosmology, and we provide details on the survey chosen to test our approach. We focus on the modelling of GW events and their distribution in \autoref{sec:theory}, while \autoref{sec:dataset} describes the method used to simulate data and the broad details of the analysis approach. We provide more details on the latter and the two specific methods used to obtain results in \autoref{sec:total_number}, where we also show the expected outcome of the methods for the survey we chose. We draw our conclusions in \autoref{sec:conclusions}.

\section{Gravitational Waves from coalescing binaries}\label{sec:gw_obs}

While different mechanisms exist to produce the GWs we observe, in this work we restrict our attention to the coalescence of compact objects, i.e. objects in binary systems that emit GWs while spiralling toward each other. The GW emission of these systems at large distances $r$ from the source can be approximated as spherical gravitational waves, hence we expect the usual $1/r$ decay. Moreover, it can be shown that the propagation through a cosmological background metric implies a decay in the amplitude as $1/d_L$, where $d_L$ is the luminosity distance of the compact binary \cite{Maggiore2007}.

Observing the waveform of an incoming GW, it is therefore possible to obtain a measurement of the luminosity distance between the observer and the progenitor system. We briefly review here the theoretical modelling of this relation, following the approach of \cite{PhysRevD.52.848,Sathyaprakash:2009xt,Zhao:2010sz}.

In the transverse traceless gauge the time domain waveform depends on the antenna patterns of the detector and on the $+,\times$ polarizations:
\begin{equation}
    h(t) = F_+(\theta, \phi, \psi)h_+(t) + F_{\times}(\theta, \phi, \psi)h_{\times}(t)\,,
\end{equation}
where $\theta$ and $\phi$ determine the position of the GW source on the celestial sphere, while $\psi$ is the polarization angle, $F_{+,\times}$ are the detector antenna patterns and $h_{+,\times}$ are the two polarizations of the wave, arising from the independent components of the metric tensor perturbation; we also include the time dependence only in these latter terms since we are considering transient sources of emission, thus neglecting the effect of modulation in the angles due to the relative motion of the detector and the source.

The two polarizations for a general inspiralling binary have the expression:
\begin{equation}
    h_{+,\times} = h_0(t) H_{+,\times}(\iota, \boldsymbol{\omega}(t))\,,
\end{equation}
with $\iota$ the inclination angle with respect to the line of sight of the binary's orbital plane, and $\boldsymbol{\omega}$ other orbital parameters, included in two orbital parameter functions $H_{+,\times}$.
As stated above, we expect the amplitude of the wave to scale with the source distance; such an effect is encoded in $h_0$, which can be written as
\begin{equation}
    h_0(t) = \frac{4 G \mathcal{M}}{c^4 d_L}(\pi G \mathcal{M}\nu_{\rm GW}(t))^{2/3}\,,
\end{equation}
where $G$ is the Newton constant, $\nu_{\rm GW}$ is the time-dependent frequency of the wave, and $\mathcal{M}$ is the chirp mass, which depends on the mass of the two binaries as
\begin{equation}\label{eq:chirpmass}
    \mathcal{M} = M\eta^{3/5}\,
\end{equation}
with $M=m_1+m_2$, the total mass of the system and $\eta$ the mass ratio defined as
\begin{equation}\label{eq:massratio}
    \eta = \frac{m_1m_2}{(m_1+m_2)^2}\,.
\end{equation}
Notice that both the GW frequency $\nu_{\rm GW}$ and the total mass $M$ are redshifted from their source value. When needed one can go to the Fourier domain from the time domain waveform
\begin{equation}
    h(f) = \int_{-\infty}^{+\infty}h(t) e^{-2\pi i f t}{\rm d}\,t\,,
\end{equation}
which can be rewritten, using the stationary phase approximation, as \cite{PhysRevD.52.848,Sathyaprakash:2009xt,Zhao:2010sz}
\begin{equation}\label{eq:f_waveform}
    h(f) = A\,f^{-\frac{7}{6}}\exp{\left[i(2\pi ft_0-\frac{\pi}{4}+2\Psi(f/2)-\varphi_{(2,0)})\right]}\,
\end{equation}
where $t_0$ is a constant describing the epoch of the merger, for simplicity set to zero in our analysis, $\Psi(f; \psi_0, \psi_i, M)$ is a phase function, depending on the phase at the epoch of the merger $\psi_0$, the post-Newtonian expansion coefficients $\psi_i$ and the total mass, and we also have an additional function of the angles $\varphi_{(2,0)}(\iota, F_{+,\times})$ \cite{Blanchet:2013haa}.

In the following, we will use the publicly available package \texttt{PyCBC13}\footnote{\href{https://pycbc.org/}{https://pycbc.org/}} to obtain the waveform $h(f)$ computation for a given event, given a set of parameters for the progenitor system, such as the binary masses or its orbital inclination. Moreover, we will be using the \texttt{IMRPhenomD} frequency domain waveform model \cite{Nitz:2018rgo,Biwer:2018osg}. 

In this work, our main goal is to extract cosmological information from GW observations. For such a reason, our main interest lies in the amplitude term $A$ of \autoref{eq:f_waveform}, where the distance of the source enters. This term can be written as \cite{PhysRevD.52.848,Sathyaprakash:2009xt,Zhao:2010sz}
\begin{equation}\label{eq:amplitude}
A = \frac{1}{d_L(z)}\sqrt{F_+^2\left[1+\cos^2{\iota}\right]^2+4F_{\times}^2\cos^2{\iota}}\sqrt{\frac{5\pi}{96}}\pi^{-7/6}\mathcal{M}^\frac{5}{6}\,.
\end{equation}

We can see in \autoref{eq:amplitude} the inverse dependence of the amplitude on the distance, but we can also notice how this amplitude depends on parameters that can vary from system to system, such as the inclination $\iota$, and the position of the source on the sky which enters in the antenna patterns $F_+$ and $F_\times$. In the following we will not consider this second class of parameters, as all the cosmological dependence lies in $d_L$, and we generate them randomly for each system (see \autoref{sec:dataset}). However, these parameters can have a significant impact on the uncertainty with which the distance can be measured; while a full treatment of their impact is left to a future work, we discuss their effect and our approximation to take them into account in \autoref{app:dl_error}.

\subsection{The Einstein Telescope}\label{sec:ET}

\begin{figure}[h]
    \centering
    \includegraphics[width=0.8\textwidth]{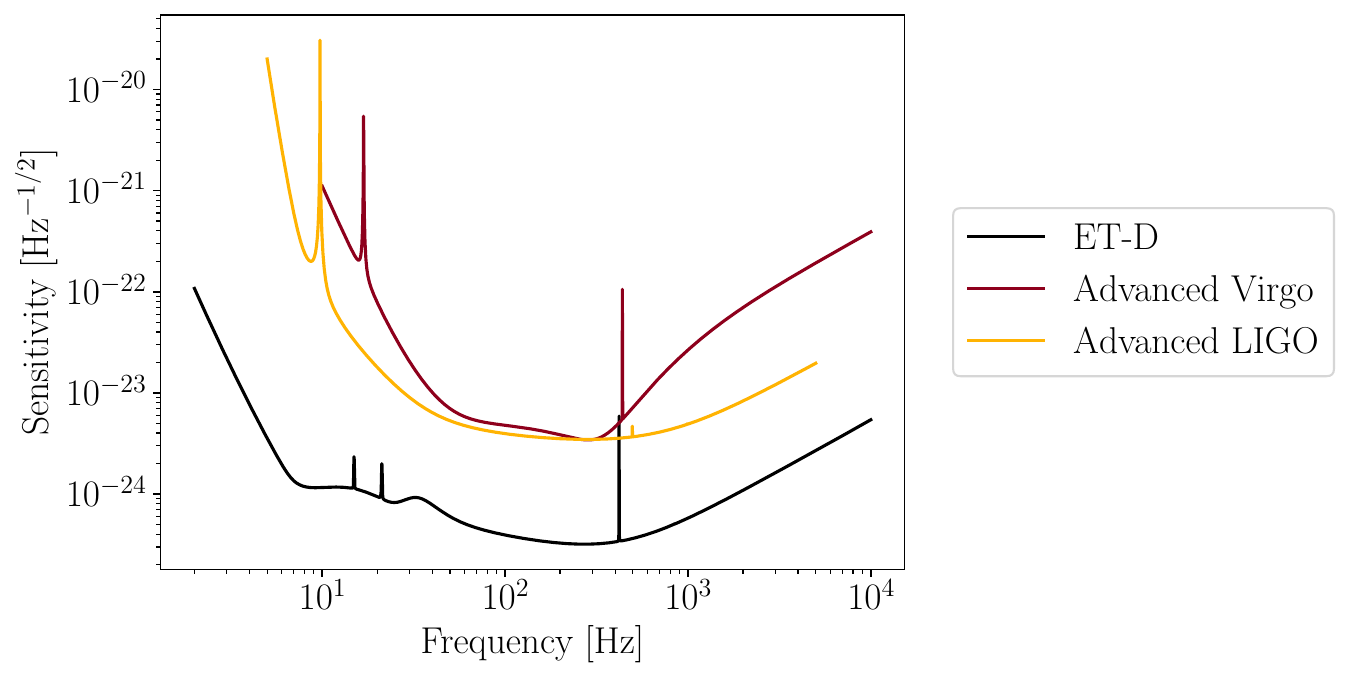}
    \caption{Sensitivity curves for Einstein Telescope interferometer in the ET-D configuration, compared to the current capabilities of Advanced LIGO and Advanced Virgo. }
    \label{fig:ET_sens_curve} 
\end{figure}

The Einstein Telescope interferometer (ET), will be part of the third generation of ground-based detectors and it is planned to start operating in the 2030s \cite{Maggiore:2019uih}. The \emph{reference design} of ET will have its arm length extended from the current Virgo 3 km and LIGO 4 km up to 10 km or even 15 km. The entire structure could be placed a few hundred meters underground and have a triangular shape for the three nested interferometers it will contain. Specifically, the detector could be composed of two different instruments, one optimized for low frequencies (ET-LF) with a low power laser and at cryogenic temperatures of 20 K; the other one for high frequencies (ET-HF) with high power and at room temperatures; thus each configuration is operated at its lowest noise condition in each frequency band. After several initial sensitivity curve models (ET-B and ET-C), the last one reached is the ET-D, which not only takes into account ET-LF and ET-HF, but also other refinements on noise models \cite{Maggiore:2019uih}; Figure \ref{fig:ET_sens_curve} shows the comparison between the noise level of current detectors (advanced Virgo and advanced LIGO) and the expected sensitivity of the ET-D design. In the mock data generation performed in this work, we will consider the ET-D sensitivity curves to extract simulated observed events.

There is also another configuration under evaluation, where two classical L shaped (2L) detectors of an extended arm length of 15 km could be placed in two different locations. The current candidates for this alternative project would be the Sardinia site of Sos Enattos; a site in the Meuse-Rhine region, or possibly in Kamenz in German Lusatia region \cite{Branchesi:2023mws}. Moreover, concerning our cosmological interest in luminosity distances, it has been shown that a 2L configuration could outperform the triangle configuration with an increased number of $d_L$ measurements with $1\%$ error, with respect to what the reference design can observe \cite{Branchesi:2023mws}.

Sticking to the reference design, we can understand the potential of third generation interferometers if we consider the maximum redshift that can be reached by the survey as a function of the total coalescing mass of a binary system: total masses of $20-100 M_{\odot}$, observed for black hole-black hole (BH-BH) or black hole-neutron star (BH-NS) binaries, will be detected by ET up to $z\sim20$ and higher, even probing the possible presence of primordial origin black hole mergers \cite{Maggiore:2019uih}. The expected detection rates will be of orders $10^5-10^6$ for BH-BH and $7\times10^4$ for NS-NS per year; moreover, the electromagnetic counterpart of NS-NS could be observed with $10^2-10^3$ events per year. With respect to the current detectors, ET will be also able to observe GW from other sources rather than just compact object binaries: the stochastic gravitational wave background, gravitational emission from isolated pulsars and from supernovae events \cite{Branchesi:2023mws}.

\section{Modelling the distribution of events}\label{sec:theory}

The main goal of this work is to exploit GW observations to obtain information on the cosmological model. This is straightforward when a GW event is observed with an electromagnetic counterpart, and therefore its redshift can be measured, as one could fit a theoretical relation between luminosity distance $d_L$ and redshift to the observed quantities (see e.g. \cite{Cai:2016sby,Du:2018tia,Hogg:2020ktc}). If we focus on Dark Sirens, instead only the measurement of $d_L$ is available, and the only observable we can exploit is the distribution of observed events in luminosity distance. While this will make extracting cosmological information more complicated, we expect the great majority of the events observed to fall in this category; developing methods that can work without any redshift measurement is therefore crucial to fully exploit the GW catalogues that will be available.

In this section we review the essential details of a widespread modelling of the GW event distribution \cite{LIGOScientific:2016ebi,LIGOScientific:2017zid}, focusing our attention on Astrophysical Black Holes (ABH), showing how this can be connected to cosmology.

It is important to stress that in this work we neglect the impact of other possible progenitors, such as Neutron Stars (NS) or Primordial Black holes (PBH). These do not follow the same theoretical distribution that we will obtain for ABH and therefore the presence of such events in an observational catalogue could affect the results. One expects however that events originating from ABH will be the dominant population \cite{KAGRA:2021duu}; for such a reason we assume in this work that ABH is the only population present, and we leave the study of the impact of NS and PBH for a future investigation.

\subsection{Distribution of astrophysical black hole binaries in luminosity distance}\label{sec:redshift_distribution}

When dealing with a GW survey, what is obtained from observations is a catalogue of events, each with their observed features, which include the event position in the sky, mass of progenitors and luminosity distance (see \autoref{sec:gw_obs}).

In order to extract cosmological information from such events, we need to compare what we observe with a theoretical prediction from a model that depends on cosmological parameters. A first feature of GW observations that we can connect to the cosmological model is the rate of events in time, as this will depend, other than on the astrophysical phenomena responsible of the creation of GW progenitors, on the rate of expansion of the Universe.

Indeed, what the predictions from cosmological models give us is the \emph{merger rate density} $\mathcal{R}(z)$, which is the number of merger events per unit of comoving volume $V_c$, per unit of proper time of the source $t_s$, as a function of redshift $z$: 
\begin{equation}
\mathcal{R}(z) = \frac{dN}{dt_sdV_c}\,.
\label{eq:MRD_def}
\end{equation}

To consider a real detection we need to change the used reference frame, moving from the source to the detector rest frame and its proper time $t_{d}$. In addition to this, we also want to obtain the probability of an event in units of redshift, a quantity that is (in principle) observable, rather than in terms of the comoving volume.
Therefore, we define the \emph{merger rate} $R(z)$, used throughout this work, as the number of merger events per unit proper time of the detector and per unit redshift, which can be obtained from the merger rate density through a change of variables
\begin{equation}
R(z) = \frac{dN}{dt_{d} dz}= \frac{1}{1+z} \frac{dV_c}{dz} \mathcal{R}(z)\,,
\label{eq:MR_MRD_relation}
\end{equation}
where we used the relation $dt_{d} = dt_s(1+z)$. 

We can see how the cosmological model enters in this transformation, as the relation of the comoving volume with the redshift leads to

\begin{equation}
\frac{dV_c}{dz} = \frac{4\pi c }{H(z|\vec{\theta}\,)} \frac{d_{L}^2(z|\vec{\theta}\,)}{(1+z)^2}\,,
\label{eq:ComVol_derivative}
\end{equation}
where $d_{L}(z|\vec{\theta}\,)$ is the luminosity distance and $H(z|\vec{\theta}\,)$ the Hubble parameter, and we explicitly show that these involve a cosmological model with their dependence on a set of cosmological parameters $\vec{\theta}$.

We finally have the expression for the merger rate function:
\begin{equation}
R(z|\vec{\theta}\,) = \frac{d_{L}^2(z|\vec{\theta}\,)}{(1+z)^3} \frac{4\pi c }{H(z|\vec{\theta}\,)}  \mathcal{R}(z)\,.
\label{eq:MR}
\end{equation}

With the merger rate $R(z)$ we can compute the total number of merger events measured during an observation time $T_{\rm obs}$ from an ideal detector:
\begin{equation}
\bar{N}_{\rm tot}(\vec{\theta}\,) = T_{\rm obs} \int_{z_{\rm min}}^{z_{\rm max}} R(z'|\vec{\theta}\,) {\rm d}\,z'\,,
\label{eq:Ntot}
\end{equation}
where events are observed only in a finite interval of redshift $z\in[z_{\rm min},z_{\rm max}]$. 

With the total number of events $\bar{N}_{\rm tot}(\vec{\theta}\,)$ and the number of events at redshift $z$, we can obtain the probability distribution of the mergers, and therefore of the GW events to take place, as
\begin{equation}
P(z|\vec{\theta}\,) = \frac{T_{\rm obs}R(z|\vec{\theta}\,)}{\bar{N}_{\rm tot}(\vec{\theta}\,)}= \frac{R(z|\vec{\theta}\,)}{\int_{z_{\rm min}}^{z_{\rm max}} R(z'|\vec{\theta}\,) dz'}\,.
\label{eq:p_z}
\end{equation}

Notice that, while this is the theoretical distribution for events to occur, this does not coincide with the distribution of observed events. As we will detail in the following, several of these events will be too far away for surveys to detect them efficiently and therefore they will not appear in the survey catalogues. Then, in the case of a real detection the quantities can be weighted by a detection function $f_{\rm det}(z)$, which represents the fraction of observable events for a given redshift $z$ and it depends on the detector considered. Therefore, we define the \emph{detected merger rate}:
\begin{equation}
    R_{\rm det}(z) = R(z)f_{\rm det}(z)\,.
    \label{eq:MR_det}
\end{equation}

Furthermore, as we discussed in \autoref{sec:gw_obs}, unless an electromagnetic counterpart is observed, GW surveys do not contain information on the redshift of the events. We can however transform the theoretical redshift probability into a probability in the space of luminosity distance, $T(d_L|\vec{\theta}\,)$, transformation which is itself a function of the cosmological model
\begin{equation}\label{eq:p_dl}
    T(d_L|\vec{\theta}\,) = P\left(z(d_L|\vec{\theta}\,)|\vec{\theta}\,\right)\Bigl|\frac{{d}}{{ d}\,(d_L)}\,z(d_L|\vec{\theta}\,)\Bigr|\,,
\end{equation}
where $z(d_L|\vec{\theta}\,)$ is the inverse function of the luminosity distance, which is defined as
\begin{equation}
    d_L(z|\vec{\theta}\,)=(1+z)\int_0^z{\frac{d\,z'}{H(z'|\vec{\theta}\,)}}\,.
    \label{eq:lumdist}
\end{equation}

\subsection{Merger rate density and the Star Formation Rate}\label{sec:rate_density_model}

With \autoref{eq:p_dl}, we have shown how we can obtain the theoretical distribution of events in the luminosity distance space, assuming a cosmological model that provides $d_L(z|\vec{\theta}\,)$ and $H(z|\vec{\theta}\,)$. Such a distribution depends, through \autoref{eq:MR}, on the merger rate density of the progenitors of the GW events. This is necessarily related to the evolution of the ABH through the history of the Universe, since these are the seeds of the GW events we are investigating. 

ABH can form at the latest stage of star evolution from the collapse of massive stars after their nuclear burning phase ends, while the formation of ABH binaries can happen through two channels:
a pre-existing binary system of stars evolves into a system of two ABHs, or two separate isolated ABHs form a binary system in a later phase. For both cases the expression for the merger rate density can be written as a function of time $t$ and black hole mass $m_{\rm BH}$ \cite{Dvorkin:2016wac,Mukherjee:2021ags}
\begin{equation}
\mathcal{R}(t,m_{\rm BH}) = N\int_{\Delta t_{\rm min}}^{\Delta t_{\rm max}}\mathcal{R}_{\rm birth}(t-\Delta t_{d},m_{\rm BH})P(\Delta t_{d}){\rm d}\,\Delta t_{d}\,,
\label{eq:rate_density}
\end{equation}

where $N$ is a normalizing factor, $P(\Delta t_d)$ is the distribution function of the time delay between ABH formation and merger ($\Delta t_d$), and $\mathcal{R}_{\rm birth}(t,m_{\rm BH})$ represents the birthrate of the ABH as a function of time and black hole mass. The latter can be expressed as
\begin{equation}
\mathcal{R}_{\rm birth}(t,m_{\rm BH}) = \int \psi_{\rm SFR}\left(t-\tau(m)\right)\,\phi(m)\,\delta\left(m-g^{-1}_{\rm BH}(m_{\rm BH})\right) {\rm d}\,m\,;
\label{eq:birth_rate}
\end{equation}
here $\tau(m)$ is the lifetime of a star of mass $m$, $\phi(m)$ is the Initial Mass Function (IMF) \cite{Dvorkin:2016wac}, $\psi_{\rm SFR}$ is the Star Formation Rate density (SFR), $\delta(m)$ the Dirac delta function, and $g_{\rm BH}^{-1}$ is the inverse of a function that, for each Zero Age Main Sequence star mass value (the mass of a star at the start of the hydrogen burning phase), gives its corresponding black hole mass $m_{\rm BH}$.

One can notice that in order to obtain the merger rate, it is required to know the properties of the systems, like the progenitor masses and the delay between the binary formation and the merger, with the latter depending on the details of the systems through their orbital parameters.

Here, we follow the approach of \cite{Dvorkin:2016wac}, where a distribution $P(\Delta t_d)\propto 1/\Delta t_d$ is considered, and the integral of \autoref{eq:rate_density} is taken with $\Delta t_{\rm min}=50$ Myr and $\Delta t_{\rm max}=H_0^{-1}$, an approach motivated by numerical simulation of binary BH formation \cite{Belczynski:2016obo}. Nevertheless, we also follow the approach of \cite{Martinelli:2022elq}, and account for the uncertainties in the modelling of the merger rate through the normalization factor $N$, choosing this by imposing that the predicted merger rate density is compatible with the observations reported in the Gravitational-Wave Transient Catalogs (GWTC) \cite{LIGOScientific:2020kqk,KAGRA:2021duu}.

Following \cite{Martinelli:2022elq}, we change variable from time to redshift, and we assume a monochromatic mass distribution for ABH, i.e. we assign the same mass to all ABH that will form the events we observe. This is a simplifying assumption because we expect from observations to have a distribution of masses for the components of the binary systems that produce the merger events. In this paper we chose the value for our fiducial ABH to be the peak of the observed current data, i.e. $m_{\rm ABH} = 7 M_{\odot}$. We expect that the effect of a distribution of masses and mass ratios would affect the SNR with respect to the monochromatic case, with this depending on the specific features of the system. We will consider our results to be the first approximation to this mass problem and discuss this systematic in a future work.

Furthermore, we assume that the birth rate is direcly proportional to the SFR \cite{Dvorkin:2016wac}
\begin{equation}
    \mathcal{R}_{\rm birth}(z) \propto \psi_{\rm SFR}(z)\,.
\end{equation}

The SFR can be measured through galaxy survey observations, which provide the rate of star formation at different redshifts, with the most distant ones being the most uncertain and where results can significantly depend on the type of tracer used to reconstruct the SFR \cite{Vangioni:2014axa}.

For such a reason, we decide to work with a parametric approach, expressing the SFR as \cite{Nagamine:2003bd} 
\begin{equation}
\psi_{\rm SFR}(z) = \nu \frac{a e^{b(z-z_{m})}}{a-b+be^{a(z-z_{m})}}\,,
\label{eq:SFR}
\end{equation}
where the parameters $a,b,\nu, z_{m}$, controlling the shape of the distribution, are obtained fitting the observational data, and therefore depend on the high redshift tracers used.

In the following, we will investigate the impact of the uncertainty on the SFR on the final cosmological results, choosing different possible measurements of the parameters. In \autoref{tab:SFRpars}, we show the values considered for the different SFR cases, which we label as
\begin{itemize}
    \item Baseline: a SFR fit obtained using high redshift observations of galaxies in the redshift range $z\in[8,10]$, corresponding to the \emph{Fiducial model} of \cite{Chen:2019irf};
    \item GRB: a SFR fit obtained using the Gamma-ray burst rate as a high redshift tracer \cite{Vangioni:2014axa};
    \item Madau-Dickinson (MD): another common SFR model, which uses a slightly different functional form, with \cite{Madau:2014bja}
    \begin{equation}
        \psi_{\rm SFR}^{\rm MD} = \nu\frac{(1+z)^{z_m}}{1+[(1+z)/a]^b}\,,
        \label{eq:SFR_MD}
    \end{equation}
    where we kept the same parameter names for the sake of simplicity.
\end{itemize}

In addition to this, we leave ourselves the possibility to include, in each of the SFR models considered, a high redshift contribution coming from population III stars (PopIII) \cite{Vangioni:2014axa}. This contribution is modelled following \autoref{eq:SFR}, with the parameter values reported in \autoref{tab:SFRpars}, and it is added to the SFR models as
\begin{equation}
\psi_{\rm SFR}^{i+{\rm PopIII}} = \psi_{\rm SFR}^{i} + \psi_{\rm SFR}^{\rm PopIII}\,,
\end{equation}
where $i$ indicates the type of SFR considered.

\begin{figure}[h!]
    \centering
    \includegraphics[scale = 0.55]{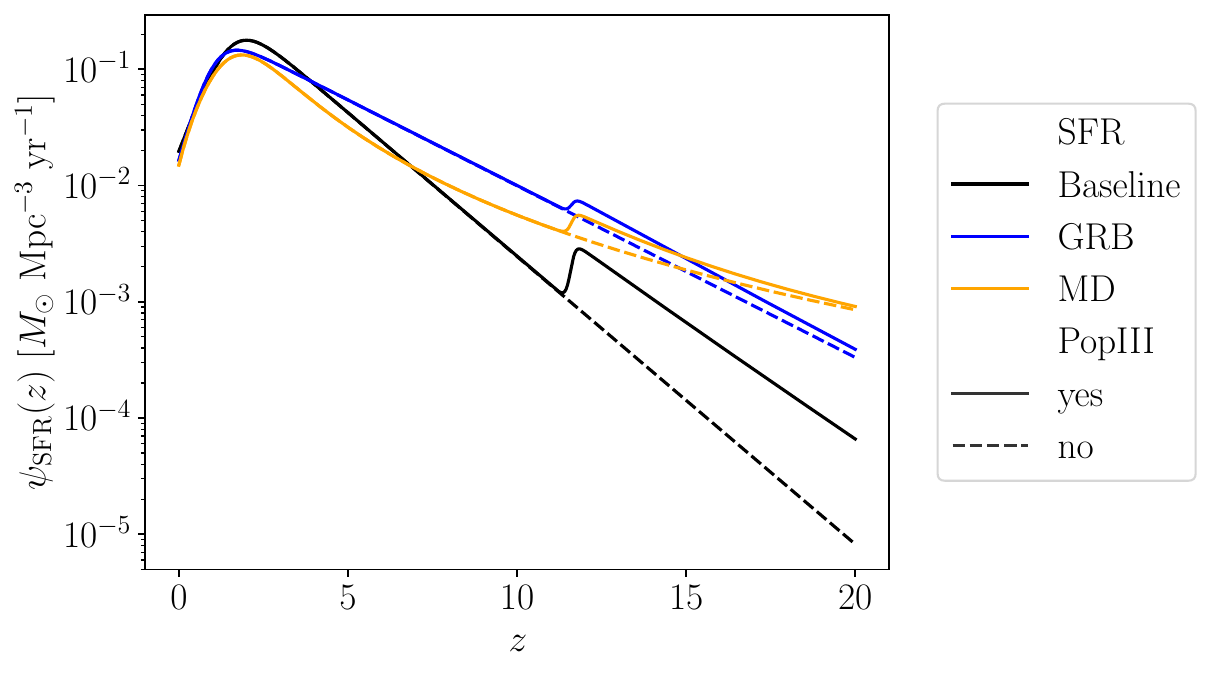}
    \caption{SFR functions for the three models, depending on the high tracers considered to reconstruct the SFR and on the presence of population III stars.}
    \label{fig:SFRmodels}
\end{figure}

The different redshift evolution of these three models are shown, both with and without the PopIII contribution, in \autoref{fig:SFRmodels}. One can notice how the main differences arise from the behaviour at high redshift, and we can therefore expect that the choice of SFR will affect the results achievable with GW surveys probing the high redshift regime. 
\begin{table}
\begin{center}
\begin{tabular}{|l|c|c|c|c|} 
 \hline
          & $\nu$   & $z_{m}$ & $a$    & $b$ \\ [0.5ex] 
 \hline
 Baseline & $0.178$ & $2.00$  & $2.37$ & $1.8$\\ 
 GRB      & $0.146$ & $1.72$  & $2.8$  & $2.46$\\
 MD       & $0.015$ & $2.7$   & $2.9$  & $5.6$\\
 PopIII   & $0.002$ & $11.87$ & $13.8$ & $13.36$\\ [1ex] 
 \hline
\end{tabular}
\end{center}
\caption{SFR parameters for the different models, depending on the high tracers considered to reconstruct the SFR and on whether or not population III stars are present.}
\label{tab:SFRpars}
\end{table}

\subsection{Impact of cosmology}\label{sec:cosmo}

As we want to extract the cosmological information from the distribution of GWs in luminosity distance, we rely on the cosmological dependence of the merger rate $R(z | \vec{\theta}\,)$, described by \autoref{eq:MR}. Such a dependence will propagate to the simulated dataset through the probability distribution of a ``true'' event $P(z |\vec{\theta}\,)$, which is connected to our theoretical model as seen in \autoref{eq:p_z}; hence, in this section we want to analyze the emerging cosmological dependence in our theoretical model $T(d_L|\vec{\theta}\,)$.

\begin{figure}[h!]
    \centering
    \includegraphics[width = 0.7\columnwidth]{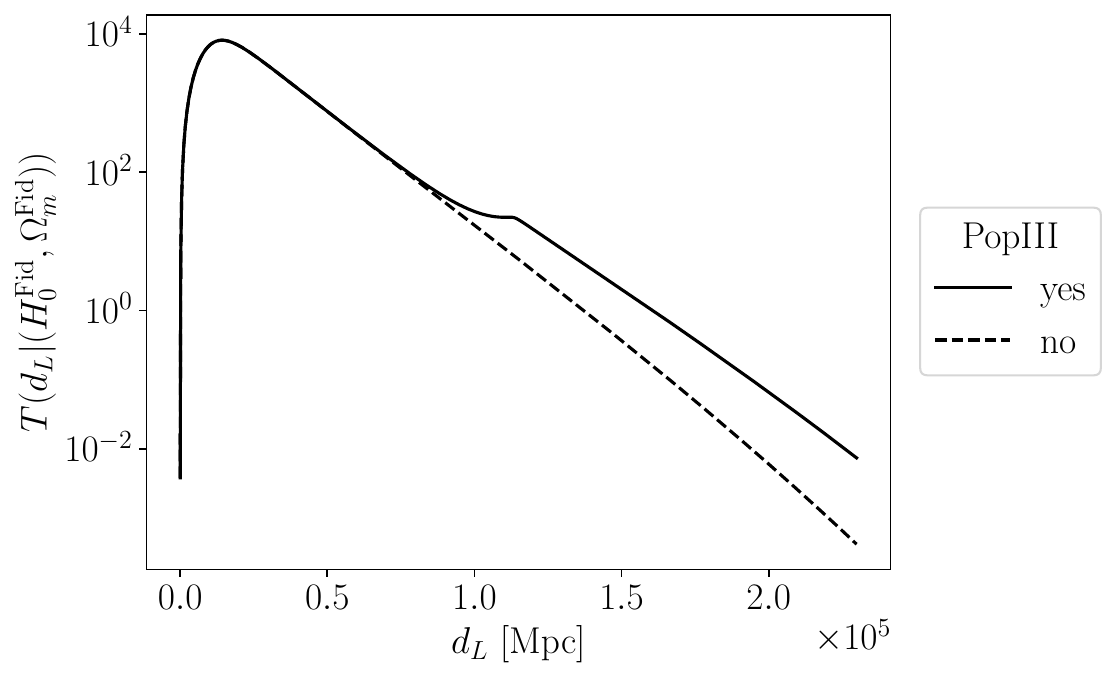}
    \caption{Theoretical distribution of merger events as a function of luminosity distance, assuming a Baseline SFR model with and without PopIII and fiducial cosmological parameters ($H_0=67$ km s$^{-1}$ Mpc$^{-1}$, $\Omega_{m}=0.32$).}
    \label{fig:T_dL}
\end{figure}

\begin{figure}[h!]
    \centering
    \begin{tabular}{cc}
    \includegraphics[width = 0.5\columnwidth] {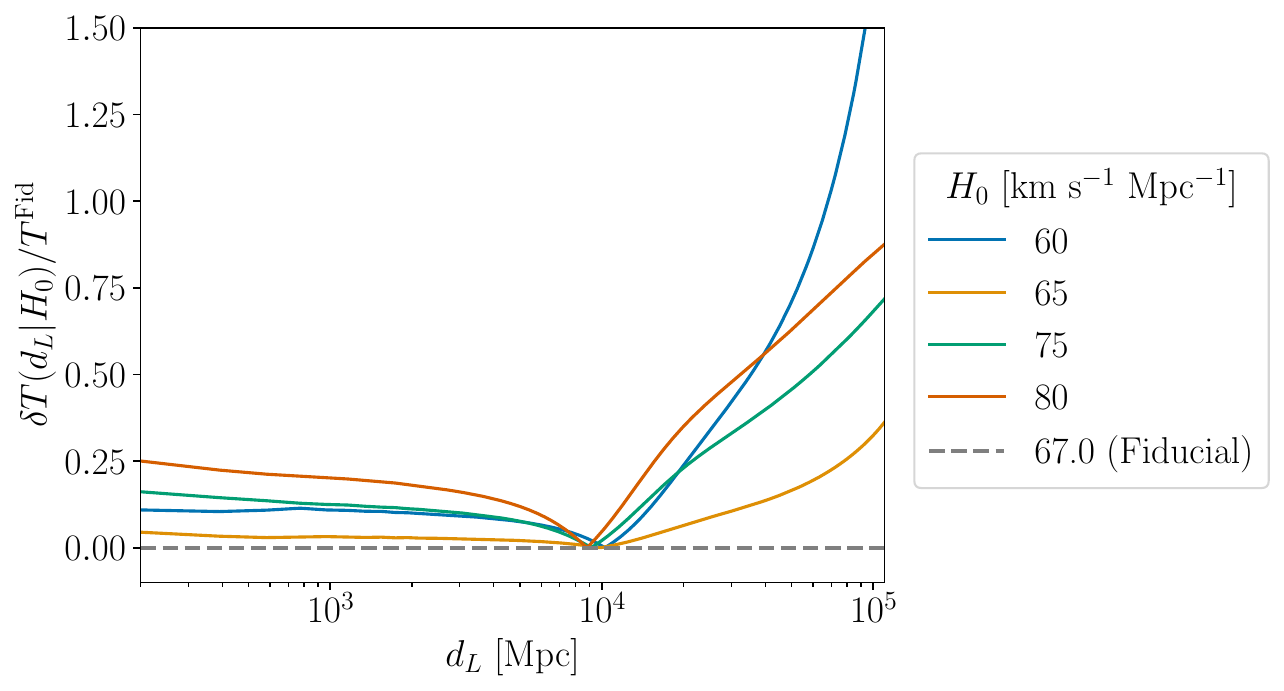} &
    \includegraphics[width = 0.5\columnwidth] {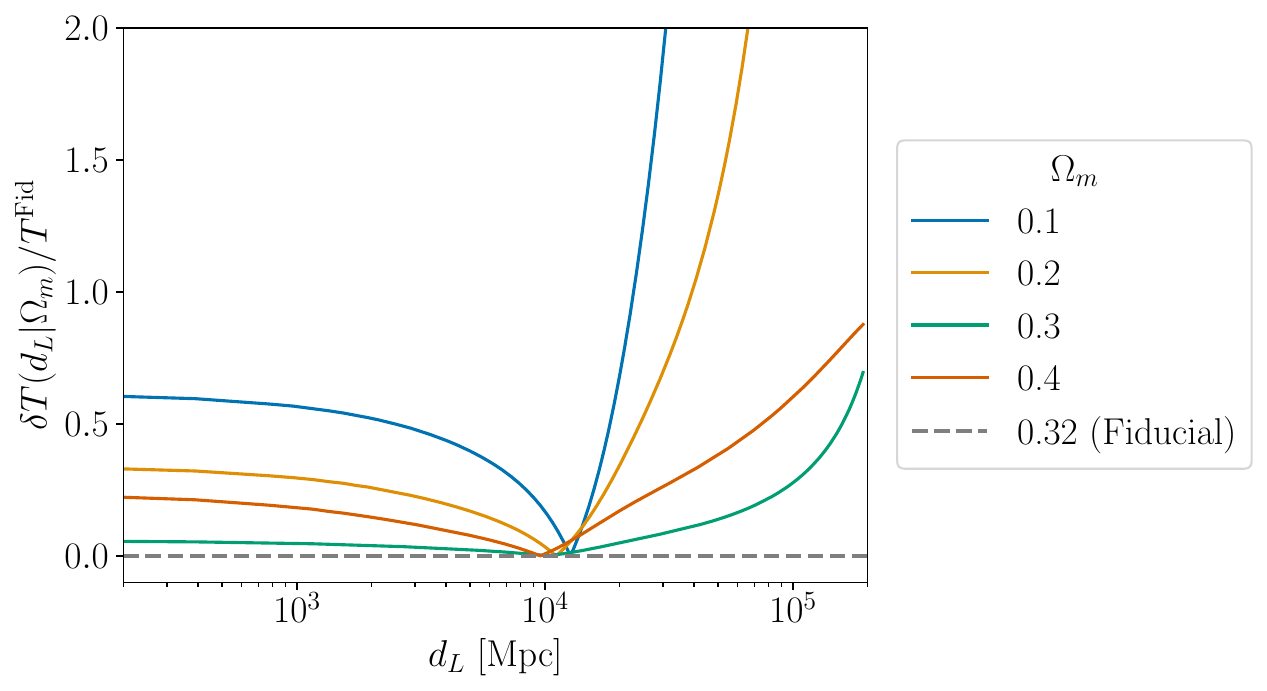}
    \end{tabular}
    \caption{Plotted here the relative absolute difference $\delta T(d_L|\theta)/T^{\rm Fid}$ of the theoretical distributions with respect to the fiducial cosmology, with a fixed Baseline SFR model (without PopIII) varying $\theta = H_0$ (left panel) and $\theta = \Omega_m$ (right panel). }
    \label{fig:cosmo_dep}
\end{figure}

To investigate this, we fix our baseline fiducial cosmology (by setting the values of $\Omega_{m}$ and $H_0$) and settle on the Baseline SFR, with the inclusion of PopIII stars, as our default astrophysical setting. Using \autoref{eq:p_dl}, we can convert the prediction on the redshift distribution of events to a distribution in distance $T(d_L|\vec{\theta}\,)$, which is the one we are able to reconstruct from data, shown in \autoref{fig:T_dL}. We can thus investigate how changing the cosmological parameters affects this distribution. In \autoref{fig:cosmo_dep}, we show in the left panel the effect of changing $H_0$, while in the right panel we vary $\Omega_{m}$.

In the fiducial distribution (\autoref{fig:T_dL}) we can observe the effect of PopIII only at high luminosity distances, where the distribution shows a small ``bump'' due to the presence of possible progenitors belonging to this stellar population. We can also notice that only at distances $\gtrsim 150$ Gpc we expect a number of events greater by an order of magnitude if PopIII stars are present; however, this happens quite far away from the peak of the distribution, where the probability of mergers is already significantly reduced, and therefore we do not expect a strong effect. For this reason, we decided to neglect PopIII when characterizing the impact of cosmological parameters in the rest of this section.

Concerning the relative differences shown in \autoref{fig:cosmo_dep}, we notice that there is generally a higher variation with respect to $\Omega_{m}$ than with $H_0$.

This cosmological dependence in the theoretical model, necessarily propagates to the total number of observed events, which is obtained from \autoref{eq:Ntot} as an integral of the distribution, including also the detection weighting of \autoref{eq:MR_det} due to a real detection. In order to investigate the cosmology dependence of this total number, we generate simulated datasets (see \autoref{sec:dataset} for more details on the steps taken for this purpose) varying the values of the cosmological parameters to observe how the total number of events changes. For each of these datasets we obtain the theoretical expectation for the total number of events $N_{\rm tot}$, and we extract the observed total number from a Poisson distribution with mean value $N_{\rm tot}$; thus, we obtain a set of values $N^{\rm tot}_i(H_0, \Omega_m)$. We then fit the trend of this observed number by varying alternately $\Omega_{m}$ or $H_0$ and fixing the other, obtaining

\begin{align}
     \ln\left(3 +\frac{\ln N}{\ln (H_0 /\rm km\, s^{-1}Mpc^{-1})}\right) &= 2.618 - 0.223 \ln\left(\frac{H_0}{\rm km\,s^{-1}Mpc^{-1}}\right)\,, \nonumber \\ \nonumber\\
     \ln\left(-\frac{\ln N}{\ln \Omega_{m}}\right) &= 1.363 + 2.510 \Omega_{m}\,,
\end{align}
which can be rewritten as power law relations

\begin{align}
     N(H_0) &= \left(\frac{H_0}{\rm km\, s^{-1}Mpc^{-1}}\right)^{\left(-3+13.705 \left(\frac{H_0}{\rm km\, s^{-1}Mpc^{-1}}\right)^{-0.223}\right)}\,, \nonumber \\ \nonumber\\
     N(\Omega_{m}) &= \Omega_{m}^{-\left(3.906 e^{2.510\Omega_{m}}\right)}
     \,.
     \label{eq:N_fits}
\end{align}

\begin{figure}[h!]
    \centering
    \includegraphics[width = 0.9\columnwidth]{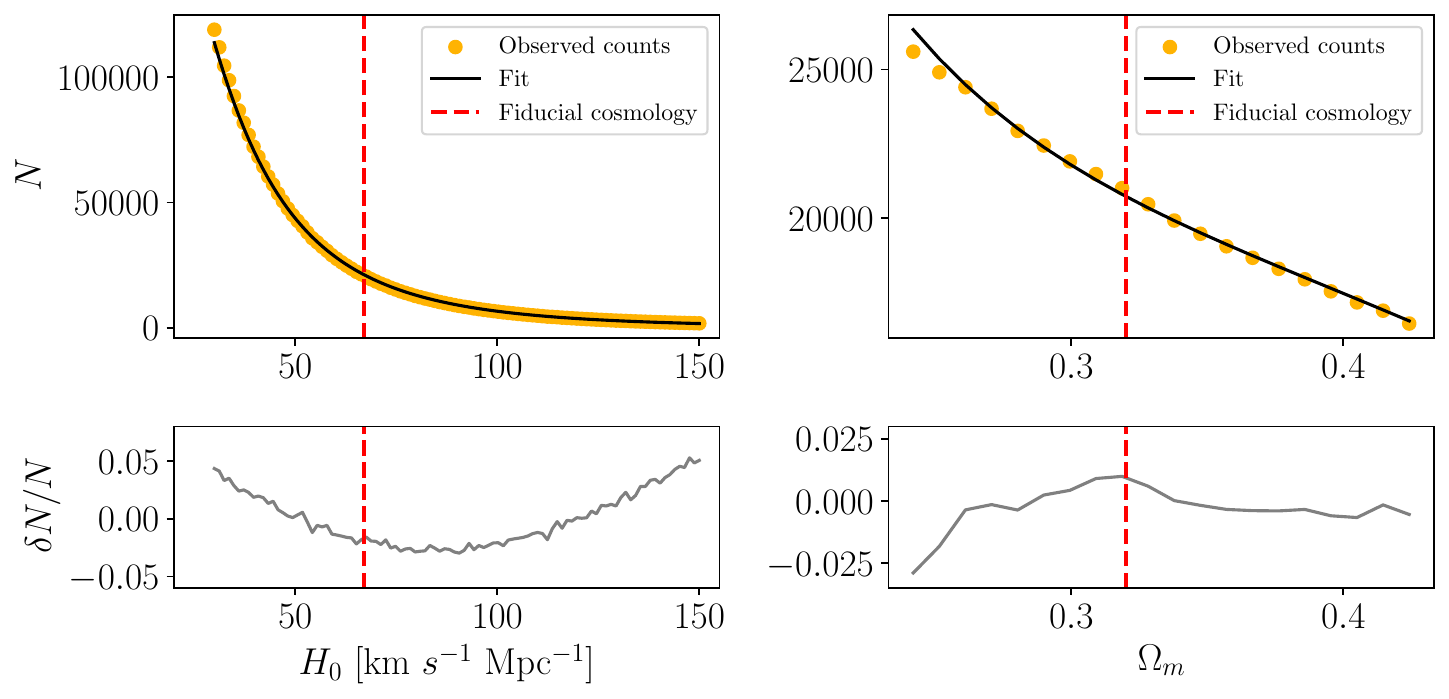}
    \caption{Comparison between the total number of the simulated datasets and those obtained with the fitting relation of \autoref{eq:N_fits}. In the upper row we have the total number of simulated observed events (without PopIII) and their respective fits. In the lower row we have the relative error between simulated observations and fits $\delta N/N$.}
    \label{fig:N_fits}
\end{figure}

In \autoref{fig:N_fits} we compare the collection of datasets with the fitted profile of \autoref{eq:N_fits} with their relative error. We notice that for both parameters the total number of events decreases with a higher value of the cosmological parameters. This is an expected result, as most of the cosmology dependence enters in the volume term of \autoref{eq:ComVol_derivative}. However, this is not the only cosmological dependence, as both the detection weighting (\autoref{eq:MR_det}), related to the observational uncertainties on the observed distances, as well as the astrophysical contribution of the merger rate density (\autoref{sec:rate_density_model}), will also change with cosmology. This leads to the deviation from the $H_0^{-3}$ dependence, that one would expect from the volume term, that we observe in \autoref{eq:N_fits}.

Overall, we find that increasing either $H_0$ or $\Omega_{m}$ decreases the number of expected events; this necessarily imply that one could obtain the same prediction with very different values of the parameters, provided that these are changed accordingly. These parameters are therefore degenerate with each other and an increase in $H_0$ could be compensated by a decrease in $\Omega_{m}$.

As we will see in \autoref{sec:total_number} this has a significant impact on the cosmological constraints one can extract from these observations. 

\section{Forecast dataset and analysis method}\label{sec:dataset}

In the previous section we have provided the details of the theoretical modelling for the distribution of ABH merging events in luminosity distance. In order to obtain information on the cosmological parameters from which such predictions depend, we want to compare our model with observations.

As pointed out in \autoref{sec:gw_obs}, current GW observations are able to probe a very limited redshift range, and therefore cannot be used for our purpose, in particular if we want to distinguish between the high redshift behaviour of  the different SFR models we introduced in \autoref{sec:rate_density_model}. For such a reason, we decide to focus our attention on the expected observations that will be performed by the Einstein Telescope, which we described in \autoref{sec:ET}.

In this section we first provide details on how we obtain a simulated catalogue for ET, and then we describe the methodology we use to compare our theoretical prediction with the simulated data to obtain cosmological constraints.

\subsection{Einstein Telescope's luminosity distance mock catalogue}\label{sec:ET_mock}

In order to generate a simulated catalogue of GW observations, we use the Python package \texttt{darksirens}\footnote{Available at \href{https://gitlab.com/matmartinelli/darksirens}{https://gitlab.com/matmartinelli/darksirens}}$^{,}$\footnote{Documentation at \href{https://darksirens.readthedocs.io/en/latest}{https://darksirens.readthedocs.io/en/latest}} \cite{Martinelli:2022elq}, which generates a catalogue of luminosity distances from merger events, given a set of observational specifications, astrophysical and cosmological parameters. This Python code was initially intended to study the fraction of Dark Matter due to PBHs, but it can be easily used without them.

The \texttt{darksirens} package requires us to assume a set of fiducial parameters to generate the simulated dataset, parameters that are divided into four categories:
\begin{itemize}
    \item \textbf{Cosmology}: the code assumes a spatially flat isotropic and homogeneous Universe with a $\Lambda$CDM cosmological background. The only cosmological parameters needed are therefore the Hubble constant $H_0$ and the total matter density parameter $\Omega_{m}$. Our dataset is generated assuming $\Omega_{m}=0.32$ and $H_0=67$ km s$^{-1}$ Mpc $^{-1}$;
    \item \textbf{Primordial Black Holes}: the $\texttt{darksirens}$ package includes the fraction of Dark Matter made up of primordial black holes $f_{\rm PBH}$ as an input parameter. As we are here assuming that all the observed events come from ABH, we set $f_{\rm PBH}=10^{-9}$, effectively neglecting the contribution of PBH;
    \item \textbf{Astrophysical Black Holes}: 
    here one should enter the parameters specifying the SFR model, together with the mass of the ABH progenitors. For the latter we follow \cite{Martinelli:2022elq} and assume that all events originate from ABH with a monochromatic mass distribution, setting $m_{\rm ABH}=7\ M_\odot$. Moreover, we produce our simulated dataset assuming the Baseline+PopIII SFR and using the corresponding parameters reported in \autoref{tab:SFRpars}. While the public version of \texttt{darksirens} supports both the Baseline and GRB models, we modified the code to have the possibility to include the MD model and to also include the PopIII contribution;
    \item \textbf{Specifications}: in order to produce a realistic dataset, \texttt{darksirens} requires the user to provide the observation time $T_{\rm obs}$, the limiting redshift to use to generate events and the threshold signal-to-noise ratio (SNR) below which an event is considered too noisy to be included in the catalogue and it is therefore discarded. We report the chosen value for the specifications in \autoref{tab:specs}. Notice that \texttt{darksirens} also allows to account for the effect of gravitational lensing on the observed distances. With respect to the investigation of \cite{Martinelli:2022elq} where the interest lied in very high redshift events, we expect a less significant impact of this effect. Neverthelss, such an effect could still be important for events at the edge of our redshift distribution. Despite this, we neglect this effect as the data simulation code only allows to take it into account for a fixed cosmology, and we will study its impact in detail in a future work.
\end{itemize}

\begin{table}[h]
\begin{center}
\begin{tabular}{ |c|c|c|c|c| }
\hline
\multicolumn{5}{ |c| }{Specifications} \\
\hline
$T_{\rm obs}$ [yrs]& SNR$_{\rm min}$ & lensing & $z_{\rm min}$ & $z_{\rm max}$ \\ 
\hline
1 & 8 & no & 0.001 & 20 \\
\hline
\end{tabular}
\caption{Simulated survey parameters and redshift interval considered. The lensing effect is neglected in this work.}
\label{tab:specs}
\end{center}
\end{table}

With all these parameters specified, we can generate the simulated dataset for ET. The process used to generate the mock is detailed in \cite{Martinelli:2022elq}, and here we report the different steps only schematically:
\begin{enumerate}
    \item the total number of mergers taking place in the limiting redshift ($N_{\rm tot}$) is computed using \autoref{eq:Ntot};
    \item for each of these events, a redshift $z_i$ is extracted from \autoref{eq:p_z} and associated to the event;
    \item using the fiducial cosmological parameters, to each redshift $z_i$ a ``true'' luminosity distance $\bar{D}_i$ is associated;
    \item for each event, the SNR $\rho_i$ is evaluated, also accounting for the event position in the sky and the orbital inclination, quantities that are randomly drawn. All events for which $\rho_i<{\rm SNR}_{\rm min}$ are removed from the catalogue, while for all others an observational error $\sigma_i = 2\bar{D}_i/\rho_i$ is computed (for more details on the computation of this uncertainty see \autoref{app:dl_error});
    \item an observational scatter $\Delta D_i$ is drawn randomly from a Gaussian distribution $\mathcal{N}(0,\sigma_i)$. The measured luminosity distance of the i-th event is then defined as $D_i \equiv \bar{D_i} + \Delta D_i$.
\end{enumerate}

After this process, a mock catalogue of measured luminosity distances with their uncertainties $\mathcal{D} = {(D_i,\sigma_i)}_{i=1,...,N_{\rm det}}$ is created, where $N_{\rm det}\leq N_{\rm tot}$ is the number of events after the SNR cut. It is important to notice that in our work we focused on the observation of Dark Sirens; this implies that, while the redshift information is computed by our simulation code, this will not be available through real observations. For such a reason, we keep in our dataset only the observed distances $D_i$ and their error $\sigma_i$, assuming this is the only information we can extract from a future ET catalogue of merger events.

\subsection{Analysis method}\label{sec:analysis}

The main focus of this paper is to quantify the constraining power of a GW survey with ET specification on cosmological parameters. Practically, this requires us to estimate the probability distribution of parameters given the data $P(\vec{\theta}\,|\mathcal{D})$. Exploiting Bayes' theorem, this can be related to the likelihood of the data $\mathcal{L}(\mathcal{D}|\vec{\theta}\,)$ as
\begin{equation}
    P(\vec{\theta}\,|\mathcal{D}) \propto \mathcal{L}(\mathcal{D}|\vec{\theta}\,) \pi(\vec{\theta}\,)\,,
    \label{eq:posterior}
\end{equation}
where $\pi(\vec{\theta}\,)$ is the prior probability of the free parameters of the analysis.

If one is able to compute $\mathcal{L}(\mathcal{D}|\vec{\theta}\,)$ given a set of parameters, it is therefore possible to estimate the posterior by sampling the parameter space and exploiting \autoref{eq:posterior}. To perform this sampling we rely on a Monte Carlo Markov Chain (MCMC) approach, making use of the Metropolis-Hastings algorithm implemented in the public code \texttt{Cobaya}\footnote{\href{https://cobaya.readthedocs.io/en/latest/}{https://cobaya.readthedocs.io/en/latest/}} \cite{Torrado:2020dgo}.

We create external modules for this code that implement the likelihood calculation we describe in \autoref{sec:total_number}, and we keep as free parameters the Hubble constant $H_0$ and the matter density $\Omega_{m}$, using flat priors with $H_0\in[60,80]$ km s$^{-1}$ Mpc$^{-1}$ when fixing $\Omega_{m}$, or $H_0\in[50,100]$ km s$^{-1}$ Mpc$^{-1}$, $\Omega_{m}\in[0.05,0.8]$, when both are free.

This approach, would in principle allow us to avoid specifying a SFR model and include the parameters for this (i.e. $\nu$, $z_m$, $a$ and $b$) as free parameters of our analysis, thus attempting to obtain information on the SFR itself from the distribution of events. However, such an analysis would be hindered by the strong degeneracies between the SFR and cosmological parameters, all entering \autoref{eq:rate_density}, and it would require the addition of external data able to provide information on the SFR parameters. This is beyond the scope of this work, and therefore we will keep the SFR parameters as fixed within our analyses, effectively assuming that such values are measured from an external survey \cite{Leandro:2021qlc}.

We will however investigate the impact of assuming the wrong SFR model: as our dataset is simulated assuming a Baseline + PopIII SFR model, we will obtain constraints on $H_0$ and $\Omega_{m}$ analysing the data with different choices of SFR to obtain the theoretical predictions. This will allow us to quantify how the constraints on cosmological parameters are affected when an incorrect modelling of the astrophysical phenomena is done.

\section{Constraints from number counts}\label{sec:total_number}

Now that we have a simulated dataset, mimicking what the ET will be able to provide, we want to assess how cosmological parameters will be constrained by it. For such a purpose, we use the approach detailed in \autoref{sec:analysis} for which we need to define a likelihood function.

In \autoref{sec:theory}, we saw how cosmological parameters enter in the calculation of the merger rate $R(z)$, both through the volume term and the merger rate density, and how this can be used to compute the distribution of events in luminosity distance, as well as their total number.

As a first step in obtaining our constraints, we only focus on the total number of observed events, i.e. we want to obtain cosmological information using as the only observable the number of events that a survey like ET will measure, neglecting any other information provided by the survey, and we leave the investigation of a distance distribution based likelihood to a future work.

\subsection{Number counts likelihood}\label{sec:num_cts_like}

We want to compare the number of events in the simulated dataset, after those below the ${\rm SNR}_{\rm min}$ are removed, with the number obtained by theoretical calculations when cosmological parameters are changed. This means obtaining the expression of the likelihood $\mathcal{L}(\mathcal{D}|\vec{\theta}\,)$ to use in \autoref{eq:posterior}, i.e. the probability of obtaining $N_{\rm obs}(\mathcal{D})$ observed events given a theoretical prediction $N_{\rm tot}(\vec{\theta}\,)$, which depends on the model parameters through \autoref{eq:Ntot}.

As shown in \cite{Martinelli:2022elq}, when considering only the total number of events, we can work with a Poissonian distribution and write the likelihood as
\begin{equation}
\mathcal{L}(\mathcal{D}|\vec{\theta}\,) = \frac{N_{\rm exp}(\vec{\theta}\,)^{N_{\rm obs}} e^{-N_{\rm exp}(\vec{\theta}\,)}}{N_{\rm obs}!}\,.
\label{eq:number_likelihood}
\end{equation}

Notice that in our definition of the likelihood we do not directly use the theoretical prediction $N_{\rm tot}(\vec{\theta}\,)$, but rather the expected number of observed events $N_{\rm exp}(\vec{\theta}\,)$. This subtle distinction is crucial, as in the simulated dataset the low SNR events are removed and this needs to be accounted for in the theoretical prediction in order to not bias the results. 

As discussed in \autoref{sec:theory}, one could define a merger rate of detected events $R_{\rm det}(z)$ and obtain all theoretical quantities from this. However, modelling the $f_{\rm det}(z)$ term entering \autoref{eq:MR_det} can be complicated, as the SNR of each GW observation depends on several features of the progenitor system.

For such a reason, we decide to obtain $N_{\rm exp}(\vec{\theta}\,)$ by generating a full mock dataset at each point of the parameter space, thus associating with each event an SNR that we can use to remove all events that are too faint from the theoretical prediction. In order to account for the dispersion in the generation of the mock dataset, we decide to average the number of observed counts over ten realizations, and to take this average as our theoretical prediction $N_{\rm exp}(\vec{\theta}\,)$.

We include this calculation in the analysis method described in \autoref{sec:analysis} and use it in the next section to obtain constraints on $H_0$, both when it is considered as the only free parameter of the analysis and when it is varied alongside $\Omega_{m}$.

\subsection{Cosmological and astrophysical constraints}\label{sec:number_constraints}

As a first result, we obtain the constraints on $H_0$ when all other parameters are fixed. While this could be interpreted as a choice of astrophysical modelling for what concerns the SFR parameters, fixing $\Omega_{m}$ implies that we are assuming it as known from some previous cosmological survey, and that constraints from GW observations are too loose to affect such a prior information.

We therefore obtain MCMC chains fitting our dataset (obtained with Baseline+PopIII) using the number count likelihood and varying $H_0$ only. Other than for the same SFR case used to obtain the dataset, we repeat the analysis also changing the astrophysical assumptions, in order to quantify the impact of such a choice on the final results.

\begin{figure}[h!]
    \centering
    \includegraphics[width = 0.8\columnwidth]{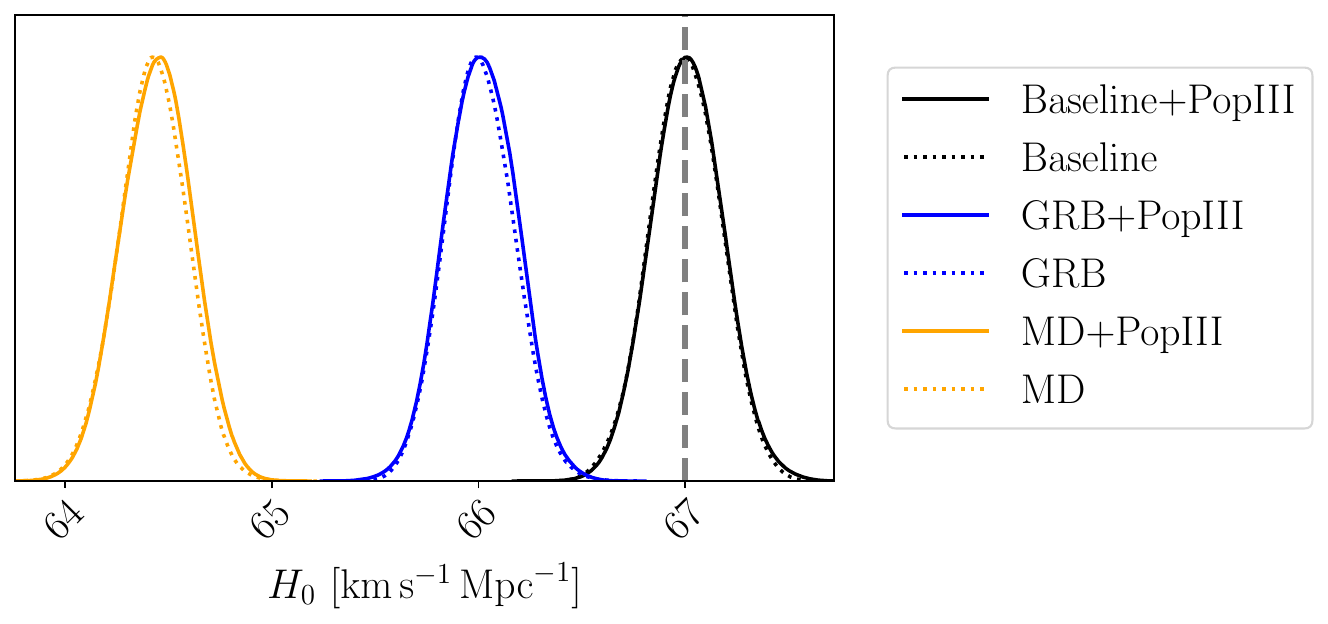}
    \caption{Posterior distribution of $H_0$ obtained analyzing the dataset constructed with a Baseline+PopIII SFR with different assumptions. Solid lines indicate the use of Baseline (black), GRB (blue) and MD (orange) in obtaining the theoretical predictions, while dashed and solid lines indicate, respectively, that the contribution of PopIII is included or neglected.}
    \label{fig:sfr_onepar}
\end{figure}

We show in \autoref{tab:results_sfr_onepar} and \autoref{fig:sfr_onepar} the results obtained in the different cases.

In the case where the analysis is performed using the same SFR as used for the dataset, we recover the value of $H_0$ used to generate the dataset, as expected, and we find a $68\%$ confidence level bound for this parameter of $H_0=67.01\pm0.17$ km s$^{-1}$ Mpc$^{-1}$. This bound is extremely tight and would be competitive with constraints obtained from the Cosmic Microwave Background (CMB) or Supernovae Ia (SNeIa). However, as we discussed in \autoref{sec:cosmo}, we expect a significant degeneracy with $\Omega_{m}$ and therefore a much looser constraint when this parameter is also free to vary.

When removing the PopIII contribution in the theoretical predictions and using these to perform the analysis, we notice no significant changes in the results, either in the recovered mean value nor in the strength of the constraint (see the Baseline entry of \autoref{tab:results_sfr_onepar}). This highlights how whether or not this effect is included does not significantly change the number of observations, a reasonable conclusion as PopIII only contributes at large redshift where the merger rate is already significantly decreasing (see \autoref{fig:SFRmodels}).

On the contrary, changing the main astrophysical assumptions and therefore switching from the Baseline to the GRB and MD star formation rates, we find no change in the error obtained on $H_0$, but a significant shift on its mean value; the results obtained assuming GRB+PopIII show a $\approx6\sigma$ deviation of the recovered $H_0$ from its fiducial value, while for MD such a shift increases to $\approx 15\sigma$. This results shows the relevance of the astrophysical assumptions in this analysis, as a wrong modelling of the SFR can lead to significantly biased results. 

We also repeat our analysis of GRB and MD cases removing the contribution of PopIII, again finding no significant changes.

The extremely large shifts obtained when changing the SFR seem to imply that this analysis could potentially distinguish between the models used. Once again, however, this result has to be considered carefully, as opening the other parameters of the analysis might significantly reduce the constraining power, and therefore hinder the possible study of SFR models.

\begin{table}
\begin{center}
\begin{tabular}{|c | c | c||} 
\hline
Assumed SFR & $H_0$ [\text{km s$^{-1}$ Mpc$^{-1}$}]\\ [0.5ex] 
\hline
Baseline+PopIII & $67.01\pm 0.17$\\ 
Baseline        & $67.00\pm 0.17$\\
GRB+PopIII      & $66.01\pm 0.17$\\
GRB             & $66.01 \pm 0.16$\\ 
MD+PopIII       & $64.44 \pm 0.17$\\ 
MD              & $64.43 \pm 0.16$\\ [0.1ex] 
\hline
\end{tabular}
\end{center}
\caption{Mean values and $68\%$ confidence level constraints obtained for $H_0$ when comparing the simulated dataset with the theoretical predictions assuming different SFR models.}
\label{tab:results_sfr_onepar}
\end{table}

Given the possible relevance of degeneracies, we now study the two parameter case, where $\vec{\theta} \equiv (H_0, \Omega_m)$. 

\begin{figure}[h]
    \centering
    \includegraphics[width = 0.8\columnwidth]{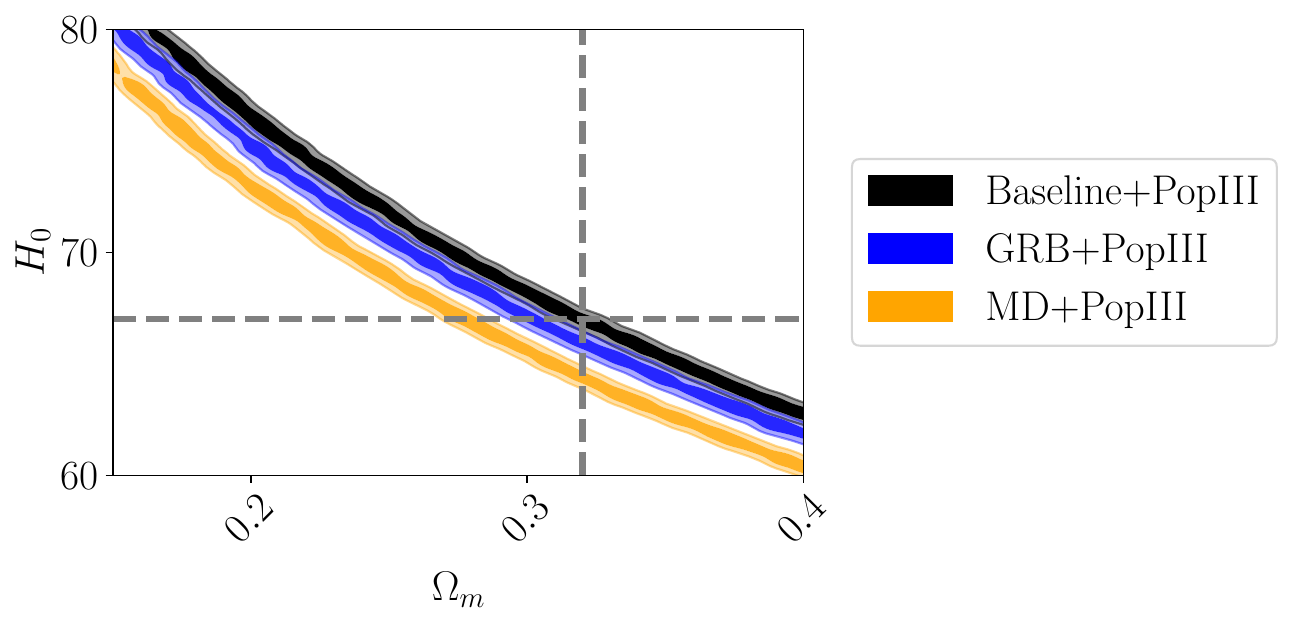}
    \caption{$68\%$ and $95\%$ confidence level contours in the $H_0-\Omega_{m}$ plane obtained from the number counts likelihood changing the SFR assumption, with black, blue and yellow referring to Baseline+PopIII, GRB+PopIII and MD+PopIII respectively.}
    \label{fig:sfr_number_twopar}
\end{figure}

We show in \autoref{fig:sfr_number_twopar} the $68\%$ and $95\%$ confidence level contours in the $H_0-\Omega_{m}$ plane, with the different contours referring to a change in the SFR assumption made in obtaining the theoretical predictions.

There is an important difference with respect to the one dimensional case; the degeneracy between $\Omega_m$ and $H_0$ implies that a change in one of the parameters can be compensated by a change in the other and this results in a complete loss of constraining power on the two parameters. As we discussed in \autoref{sec:cosmo}, only a combination of these two parameters can be constrained with the number count likelihood, but not the two separately. In order to break such a degeneracy, one needs to either include information from other surveys, e.g. adding a prior on $\Omega_{m}$, or, possibly, exploit additional information coming from GW observations, such as the distribution in distance of the events.

We also notice that the SFR model assumption is still relevant and an overall shift of the contours similar to the one dimensional case is still present, although it is less statistically significant. Indeed, GRB and MD predict more events at high redshift than the Baseline model, however more of these events will be cut from the dataset due to their low SNR, hence the total number of events for GRB and MD is lower; to account for this effect the inferred cosmological parameters should be lower, since the total number of events is inversely proportional to $H_0$ and $\Omega_m$. 

Concerning the contribution of PopIII, we also find in this case that no conclusion can be drawn concerning the presence of this population, similarly to the one parameter case. For such a reason, in \autoref{fig:sfr_number_twopar} we do not report results without the contribution of PopIII.

\subsection{Testing the SFR with the cut-and-count method}\label{sec:cut_n_count}

As we saw in the previous section, when both cosmological parameters are free to vary, the information brought by the total count of events is not enough to obtain constraints. Nevertheless, we have seen how we could still obtain some hints on the differences between the considered SFR.

In this section, we try to obtain further information on the SFR by exploiting a different method, i.e. the \emph{cut-and-count method} \cite{Martinelli:2022elq}. Indeed, we can notice from \autoref{fig:SFRmodels} that different SFR models have similar but not equal behaviours depending on redshift, especially at $z>10$ where we have the biggest differences, especially when PopIII is considered. Therefore, if we were to divide the data in two bins (left and right of a given threshold), as it is done in the \emph{cut-and-count method}, we could attempt to detect a difference in the number of counts in the right bin, depending on the SFR model, as a direct effect of the shape of these distributions; hence this method could in principle distinguish different SFRs and identify the correct one from the data.

Considering the observed luminosity distance dataset $\mathcal{D}$, we choose a value $D_*$ to cut the data into two subsets and count the number of events above the cut $N_>(\mathcal{D}, D_*|\vec{\theta}\,)$. Such a number will be affected by the parameters of the problem, i.e. cosmological and astrophysical parameters.

In accordance to the previous section we test the effect of the Hubble parameter and the assumption on the underlying SFR model, hence we construct datasets $\mathcal{D}_{H_0}^{^{\textrm{\tiny SFR}}}$ and compute the quantities $N_>(\mathcal{D}_{H_0}^{^{\textrm{\tiny SFR}}}, D_*)$. We obtain the error associated with the number above the cut  for the observed dataset $\sigma_>(\mathcal{D}, D_*)$ and for the test datasets $\sigma_>(\mathcal{D}_{H_0}^{^{\textrm{\tiny SFR}}}, D_*)$, consisting of a Poissonian term, due to the discrete occurrences of the events in the distance interval observed, and a binomial term, due to a distance's true value being above or below the cut based on its measured error \cite{Martinelli:2022elq}.

We quantify the discrepancy between the dataset and possible test datasets with the \emph{statistical shift}:
\begin{equation}
    \mathcal{S}(\mathcal{D}_{H_0}^{^{\textrm{\tiny SFR}}}, D_*) = \frac{\left|N_>(\mathcal{D}_{H_0}^{^{\textrm{\tiny SFR}}}, D_*) - N_>(\mathcal{D}, D_*)\right|}{\sqrt{\sigma_>^2(\mathcal{D}_{H_0}^{^{\textrm{\tiny SFR}}}, D_*) + \sigma_>^2(\mathcal{D}, D_*)}}\,,
\end{equation}
i.e. the distance between the test dataset, obtained with a specific SFR choice, and the observed one, in units of the error.

As a first step, we want to compare the results obtained using this method with those found in \autoref{sec:number_constraints}. Thus, we first take our threshold distance to be $D_* = 0$, because this is the case when all the events are considered. As before, we vary $H_0$ in the interval $[60, 80]$ km s$^{-1}$ Mpc$^{-1}$ and use the same mock dataset. We compute the statistical shift as a function of $H_0$ for all the SFR models considered. We expect the shift to yield the same results as above, with the minimum of the function lying in the same $H_0$ values of \autoref{tab:results_sfr_onepar}.

\begin{figure}[h]
    \centering
    \includegraphics[width = 0.8\columnwidth]{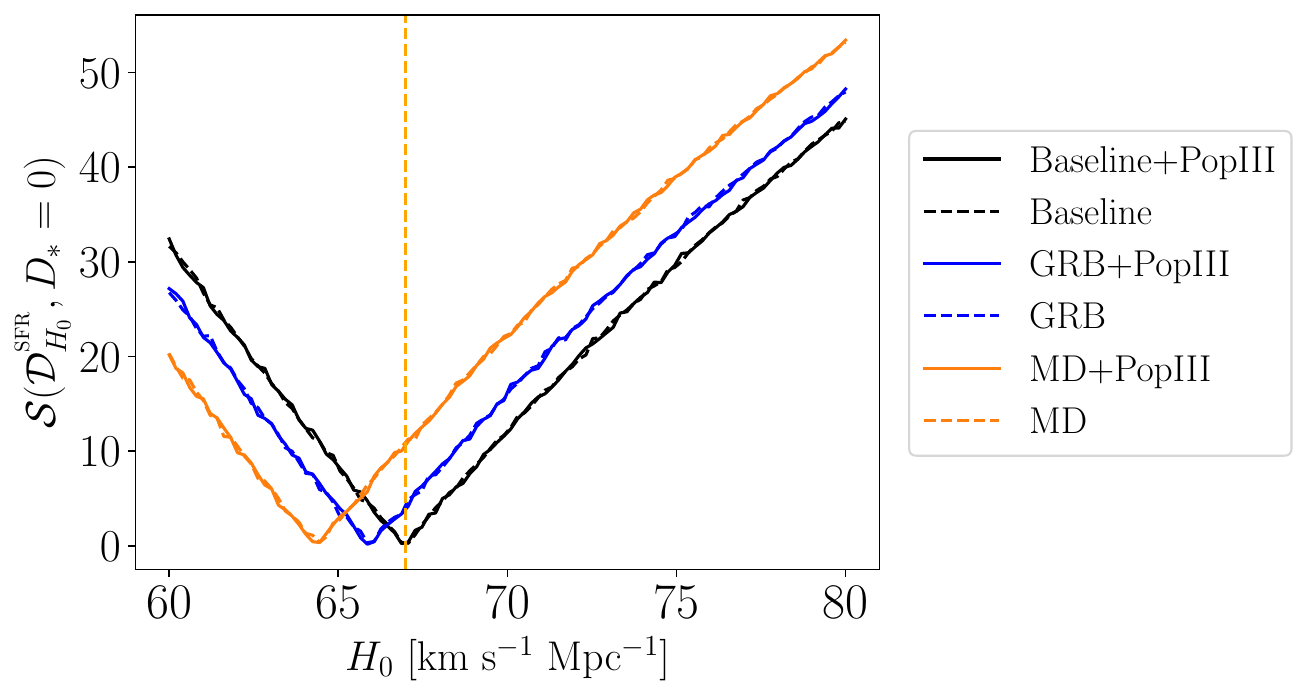}
    \caption{Statistical shift without cutting data ($D_* = 0$) as a function of the Hubble parameter for the Star Formation Rate models considered. We can observe a statistically significant shift from the fiducial cosmological value (orange dashed line) similarly to \autoref{fig:sfr_onepar}.}
    \label{fig:significance}
\end{figure}

Indeed, we can notice from \autoref{fig:significance} that the minimum of each curve, being the best estimation for each SFR model, has the same bias found with the previous method in \autoref{fig:sfr_onepar} and we still cannot distinguish the effect of PopIII stars.

\begin{figure}[h]
    \centering
    \includegraphics[width = 0.85\columnwidth]{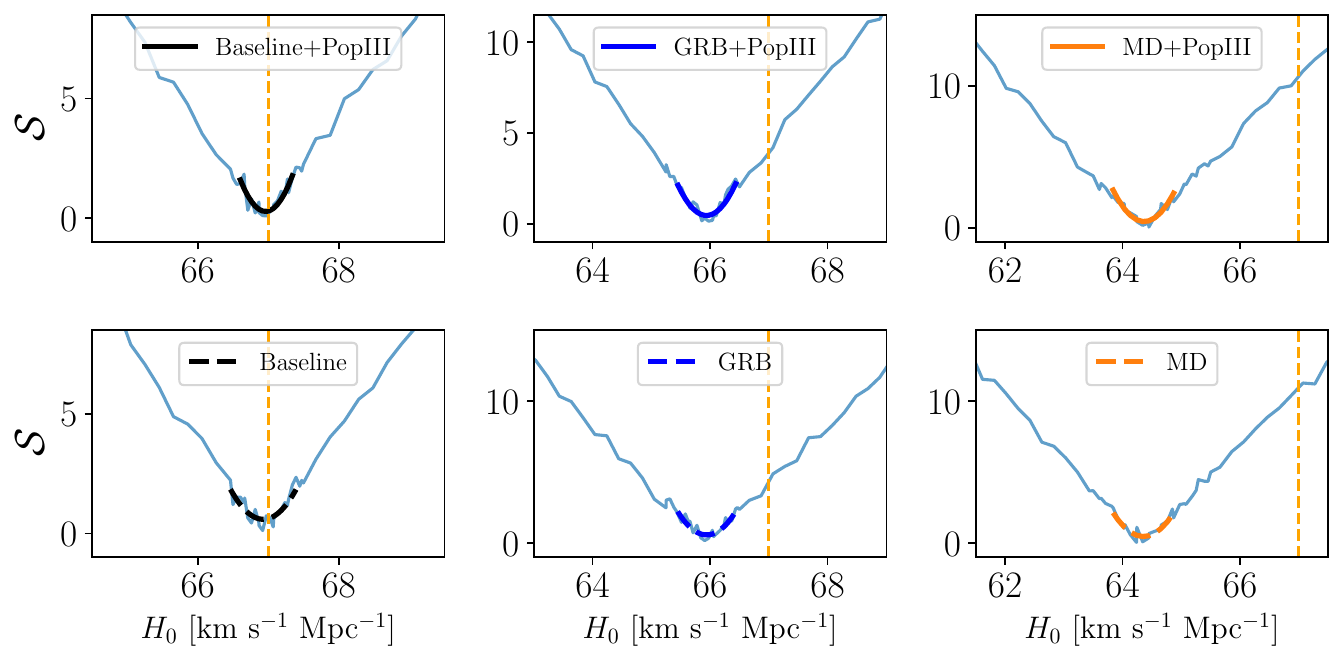}
    \caption{Detail of each statistical shift minimum with its respective fit, used to give this method's estimate of the Hubble parameter, where the fiducial cosmological value is represented by the orange dashed line}
    \label{fig:significance2}
\end{figure}

We further verified the validity of this method obtaining a qualitative estimate of $H_0$, by fitting around the minimum of each significance curve (see \autoref{fig:significance2}), then shifted the whole fit by the minimum's value and took 1$\sigma$ errors, i.e. where the shifted fit equals 1. The results are presented in \autoref{tab:sign_table} and they agree with our initial results shown in \autoref{tab:results_sfr_onepar}.

\begin{table}
\begin{center}
\begin{tabular}{|c | c | c|} 
\hline
Analysis Model & $H_0$ [\text{km s$^{-1}$ Mpc$^{-1}$}]\\ [0.5ex] 
\hline 
Baseline+PopIII & $66.96\pm 0.31$\\ 
Baseline & $66.93\pm0.41$\\
GRB+PopIII & $65.94\pm0.37$\\
GRB & $65.94\pm0.38$\\ 
MD+PopIII & $64.37\pm0.35$\\ 
MD & $64.36\pm0.40$\\ [0.1ex] 
\hline
\end{tabular}
\end{center}
\caption{Summary of the results of \autoref{fig:significance2} for different SFR models for the cut-and-count method.}
\label{tab:sign_table}
\end{table}

We now apply the cut-and-count method to investigate if this approach could be used to distinguish between the SFR models. In this case we assume the cosmological parameters as fixed to the fiducial cosmology.

Since the statistical shift is a measurement of the deviation of the number of events, having a different choice of SFR could yield different counts above the chosen threshold; this is due to the different behaviour of the SFR functions at high redshifts (see \autoref{fig:SFRmodels}).

When choosing the cut at $D_* = 0$ we cannot identify the differences in the shape of the SFR functions, only on the overall effect is has on the total counts. However, if we vary the cut across the redshift interval of observations $z_*$ (directly linked to the luminosity distance interval, because we are fixing the cosmological parameters, thus $z_*$ and $D_*$ are interchangeable) we can better highlight the difference in the SFR functions, and thus also notice where two SFR models could produce the same results.

\begin{figure}[h]
    \centering
    \includegraphics[width = 0.8\columnwidth]{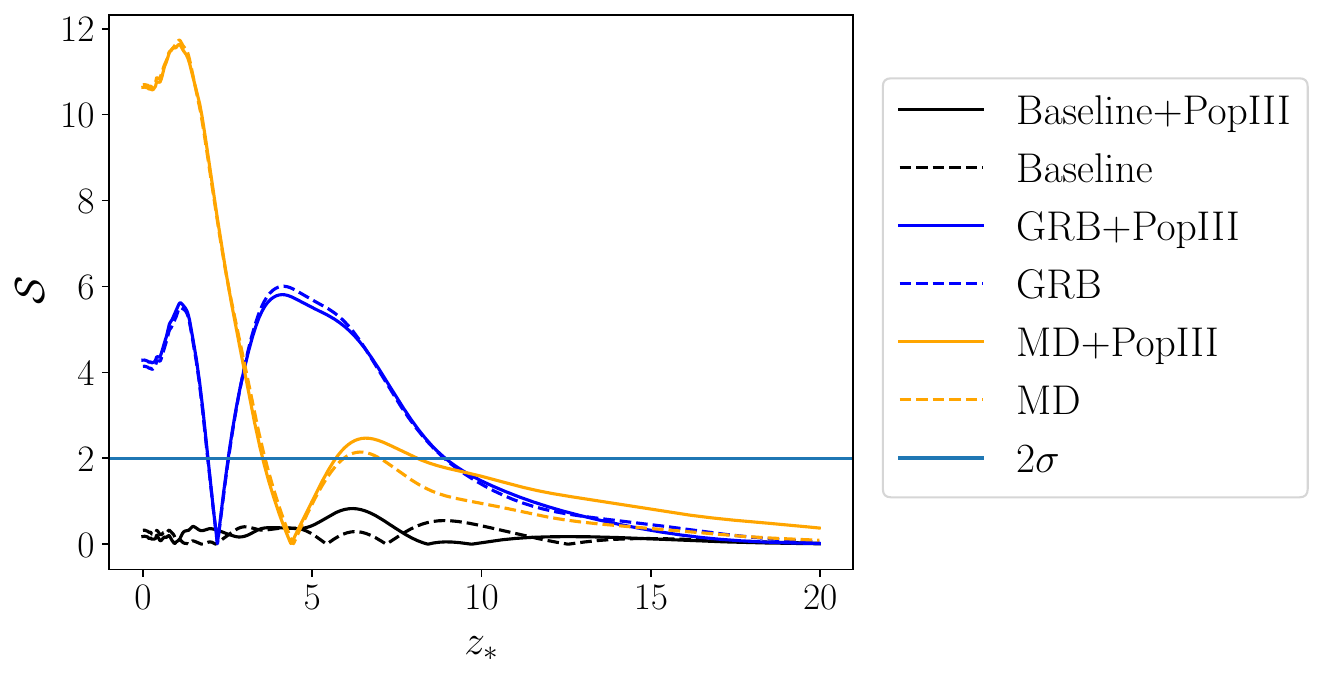}
    \caption{Statistical shift as a function of the cut in redshift for different SFR models. Here the cosmological parameters are fixed at their fiducial values.}
    \label{fig:significance_z}
\end{figure}

We computed the statistical shift at fixed cosmology as a function of the redshift cut for all the SFR models. We can see the results in \autoref{fig:significance_z} from which we can understand the following:
\begin{itemize}
    \item when the SFR model considered for the analysis coincides with that of data, $\mathcal{S}$ takes values of at most $1\sigma$, pointing out that we recover the correct model. However, with this method we are still not able to distinguish the presence of PopIII stars;
    
    \item if we consider the value of $\mathcal{S}$ in $z_* = 0$ for different SFR models, we recover the same bias as shown in \autoref{fig:sfr_onepar}, but this bias gets even larger when varying $z_*$, reaching its maximum at the first peak in $\mathcal{S}(z_*)$. The value of $z_*$ where we reach this first maximum is therefore the ideal cut to perform on the data if one is interested in observing their high redshift behaviour and studying the impact of the assumptions made on the SFR. We also notice from \autoref{fig:significance_z} how this $z_*$ is roughly the same for both the alternative SFR functions studied here ($z_*\approx2$), possibly showing that such a choice is the most suited to study this effect;
    
    \item At the respective first minimums of $\mathcal{S}(z_*)$ for GRB and MD in \autoref{fig:significance_z}, we expect a degeneracy between our fiducial model and the others, meaning that for a certain cut, indicated as $\tilde{z}_*$, we have the same number of events after it, thus they should also predict the same estimate for $H_0$ and minimizing the bias.
    Such a value of $\tilde{z}_*$ will coincide with the cross-over point of the different SFR functions, i.e. the redshift at which the alternative model considered starts predicting a higher probability to see merger events with respect to the Baseline, as it can be seen in \autoref{fig:SFRmodels}.
    
    For higher cuts than $\tilde{z}_*$, such a degeneracy will be broken and the statistical shift increases once again, until the redshift becomes high enough that the low number of expected events becomes so low that the significance of the shift vanishes.
\end{itemize}

This analysis shows that, if cosmology is assumed to be given by some external data, one could implement the cut-and-count method; changing the SFR and applying different $z_*$ cuts would allow to understand whether the SFR requires a shift in cosmological parameters in order to match the observed counts (i.e. when we find a maximum in the $\mathcal{S}$ function), or rather it is able to reproduce observations (low and flat $\mathcal{S}$).

We want to stress that, in the context of a real GW survey, e.g. with ET, this method could be relevant in extracting information on the astrophysical model simply by cutting the data above a specific threshold, as prescribed by our analysis. Then, counting the number of events above this threshold would allow to find early results faster than a parameter space sampling method.

\section{Summary and outlook}\label{sec:conclusions}

With the continuous improvement of GW observations and of the quality of the catalogue provided by observational surveys, it is becoming timely to investigate how this new window on the Universe can be used to extract cosmological information.

This is straightforward in the presence of an electromagnetic counterpart to the observed GW or when other means to measure its redshift are available. However, the vast majority of GW events we will observe do not provide this information.

Given the extremely high number of events we expect to observe with next generation surveys, it is crucial to find analysis methods able to extract cosmological information even from Dark Sirens, i.e. events for which no measurement of their redshift is available.

In this paper we focused on this class of events, assuming that they are all generated from mergers of ABHs. This assumption allows us to model the merger rate as seen in \autoref{sec:redshift_distribution} and \autoref{sec:rate_density_model}, depending on their unknown redshift, from which we can obtain the theoretical luminosity distance distribution $T(d_L)$. Notice that ABH are expected to be the dominant population of GW progenitors; assessing the impact of additional progenitor populations, such as NS and PBH, is however important to obtain accurate results, and we leave this investigation for a future work.

In \autoref{sec:cosmo} we discussed the dependence of the theoretical merger rate on two cosmological parameters: $H_0$ and $\Omega_{m}$, assuming a flat $\Lambda$CDM cosmological model. By changing the distribution in distance of the events and, consequently, the capability of detecting them with GW surveys, the final observed effect is a dependence of the total number of observed events on these two parameters, which can be exploited for cosmological inference.

We performed two analysis methods: a likelihood-based method presented in \autoref{sec:num_cts_like}, and the cut-and-count method defined in \autoref{sec:cut_n_count}, both adapted from \cite{Martinelli:2022elq}. We applied these methods on a simulated dataset obtained using the $\texttt{darksirens}$ package, considering specifications that mimic those expected for the Einstein Telescope interferometer. In producing our simulated dataset, we assumed a fiducial cosmology specified by the parameters $H_0 = 67$ km s$^{-1}$ Mpc$^{-1}$ and $\Omega_{m} = 0.32$, and assumed the Baseline+PopIII model to describe the star formation rate (see \autoref{sec:ET_mock}).

In our likelihood approach, we focused on the number counts of GW events as the observable from which to extract cosmological information. 
We defined the likelihood as a Poisson distribution on the expected number of events, and used this to constrain the free cosmological parameters. We first assumed $\Omega_{m}$ as known, possibly from some other cosmological survey, and constrained $H_0$ alone. The results for the posterior of $H_0$ are presented in \autoref{fig:sfr_onepar} for different choices of SFR functions and the respective estimates with 68\% credible intervals are presented in \autoref{tab:results_sfr_onepar}. 

The results in \autoref{fig:sfr_onepar} show that Dark Siren observations from ET could constrain the Hubble constant at a sub-percent level from the number counts of mergers. This is a very tight constraint, extremely competitive with those currently known from CMB \cite{Planck:2018vyg} or SNeIa analyses \cite{Riess:2021jrx}. We must however point out that these sub-percent estimates are obtained without considering other possible sources of error, such as the degeneracy with $\Omega_{m}$ that we highlighted in \autoref{sec:cosmo}, or systematic effects. Indeed, we assume the SFR parameters as perfectly known, and we also consider a monochromatic mass distribution for the ABH population. Moreover, we do not include further possible population that would need a different modeling, such as Neutron Stars or, possibly, Primordial Black Holes \cite{Martinelli:2022elq}.

We also explored the dependence of the results on the astrophysical assumptions done, assuming in this case a perfectly known cosmology, namely on the choice of the SFR function applied. We consider three different cases, i.e. the Baseline, the GRB, and MD models (see \autoref{sec:rate_density_model}) which differ by their high redshift behaviour. We also considered the impact of the possible presence of population III stars as progenitors of the GW systems, including their distribution in the predictions done.

We assumed the Baseline+PopIII case in building our simulated dataset and investigated the impact of an incorrect choice of the SFR when obtaining results. We found that although we can distinguish between the three SFR models, by observing a statistically significant shift of the distribution, the same cannot be said for the presence of population III stars. Indeed, in all cases, the absence of population III stars does not result in an appreciable difference from the expected values of the parameters. Thus, we do not expect the number counts of Dark Sirens to identify the presence of this stellar population.

We also relaxed our assumptions and tried to constrain both $H_0$ and $\Omega_{m}$ in \autoref{fig:sfr_number_twopar}, thus finding an expected degeneracy due to the luminosity distance expression. This result highlights how, in order for GW surveys to be competitive in their cosmological constraints, additional information is needed, either coming from external surveys, e.g. with priors on the matter content of the Universe, or by exploiting further the GW survey itself, e.g. including in the analysis also the distribution of the events in distance, rather than only relying on the total counts. This applies in particular if one wants to constrain SFR parameters which would add an extra degeneracy in the merger rate of \autoref{eq:MR} on top of cosmological parameters, especially without well defined priors for the whole set of parameters.

When applying the cut-and-count method, we obtain the theoretical prediction for a given SFR \footnote{The cut-and-count method only uses number counts, which we saw suffer from degeneracies with SFR parameters. For such a reason we fix these, and perform multiple analyses assuming the different SFR models.} on the total number counts after a redshift (distance) threshold $z_*$ ($D_*$), and compute the statistical shift with respect to the observations. Even though it is simple in its approach, we found consistent results with the likelihood-based method when $z_*=0$, i.e. where the results of the two methods should coincide. With this approach we studied different possible cuts and identified for which of these we maximize the statistical shift when a wrong SFR model is chosen. Such a result can be helpful in designing a strategy to use GW surveys to probe the SFR, as it provides insight on the subset of data containing the best information for this purpose. 

In conclusion, we found that the sensitivity of Einstein Telescope will allow us to use its observations both to constrain cosmological parameters and to infer precious details on the astrophysical assumptions done in modelling the progenitors of the GWs we observe.

While our work can be considered as a proof of concept for the methods considered, we must point out that, in order to obtain more realistic constraints, several of the assumptions made here must be relaxed, mainly allowing for more general cosmologies (e.g. including curvature or evolving dark energy), allowing for a non-monochromatic mass distribution of the binary Black Holes, and including possible systematic effects, such as the presence of other progenitor populations that will affect the theoretical predictions for the merger rate.

Furthermore, we have highlighted how in a case where no external cosmological information is included, the number counts of GW events suffer from strong degeneracies between the cosmological parameters. It is therefore necessary to include further information coming from GW surveys, such as that coming from the statistical distribution of events in distance or from a subset of events for which redshift can be measured.

\appendix 
\section{Uncertainties on luminosity distance measurements}\label{app:dl_error}

\begin{figure}[h!]
    \centering
    \includegraphics[width = 0.8\columnwidth]{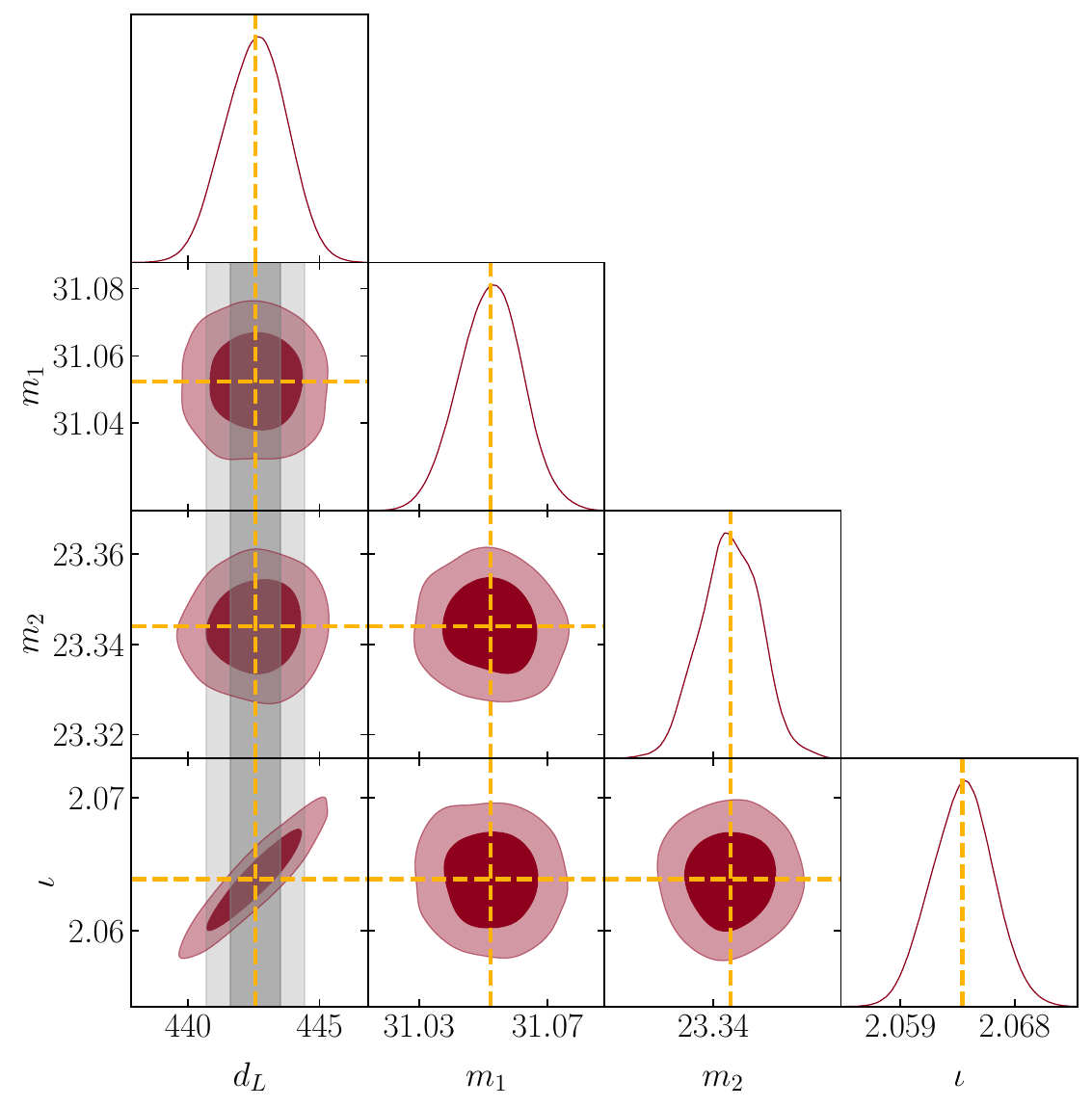}
    \caption{Example of Fisher matrix based uncertainties on the parameters inferred from a GW observation when the luminosity distance is considered alongside the masses and the inclination of the system (red contours), compared with the approximated error on the luminosity distance obtained from \autoref{eq:dl_error} (gray band).}
    \label{fig:fisher}
\end{figure}

The dataset we obtain from the \texttt{darksirens} package contains an estimate of the observational uncertainties on the luminosity distance measurements obtained as 
\begin{equation}\label{eq:dl_error}
    \sigma_i = 2\,\frac{d_L(z_i)}{\rho_i}\,,
\end{equation}
where $\rho_i$, the SNR ratio of the $i^{th}$ event is obtained as $\rho_i=\sqrt{\langle h|h \rangle}$, with $h$ the waveform in Fourier space we defined in \autoref{eq:f_waveform} and the inner product of two generic functions $h_1$ and $h_2$, $\langle h_1|h_2 \rangle$, defined as 
\begin{equation}\label{eq:inner_product}
    \langle h_1|h_2 \rangle = 4\,{\rm Re}\left[\int_{f_{\rm min}}^{f_{\rm max}}{\frac{h_1(f)\,h_2^*(f)}{S(f)}}\right]\,
\end{equation}
with $f_{\rm min}$ and $f_{\rm max}$ the limiting frequency of the instrument's sensitivity range, and $S(f)$ the noise power spectral density.

It can be shown that for a GW survey, the Fisher information matrix can be written as \cite{Vallisneri:2007ev,Gair:2022fsj} 
\begin{equation}\label{eq:fishmat}
    F_{\alpha\beta} = \left\langle\frac{\partial\,h}{\partial\,\theta_\alpha}|\frac{\partial\,h}{\partial\,\theta_\beta}\right\rangle\,,
\end{equation}
with the $\theta_\alpha$ being all the observable quantities entering \autoref{eq:f_waveform}.

If one considers only the luminosity distance $d_L$ entry of the Fisher matrix, and considering that $\partial h/\partial d_L=-h/d_L$, one obtains 
\begin{equation}
    F_{d_Ld_L} = \frac{\langle h|h \rangle}{d_L^2}=\frac{\rho_i^2}{d_L^2}\,.
\end{equation}

Considering all other observables as fixed, the error on the luminosity distance can be obtained from the square root of the inverse Fisher matrix 
\begin{equation}
    \sigma_i = \frac{d_L}{\rho_i}\,.
\end{equation}
This however, does not include the degeneracy of the luminosity distance with other observables of the waveform, which unavoidably increases the observational error on $d_L$. To account for this a factor of two is included in \autoref{eq:dl_error}, boosting the uncertainty on $d_L$ to roughly account for the extra contribution to the error coming from other uncertainties. In \autoref{fig:fisher} we show, for a single event, how the error predicted by \autoref{eq:dl_error} is comparable with the one given by a Fisher matrix obtained accounting for the uncertainties not only on $d_L$, but also on the two masses of the system ($m_1$ and $m_2$) and on its inclination ($\iota$). Given this result, we rely in the production of our dataset on the approximated results of \autoref{eq:dl_error}, rather than performing a full Fisher analysis for all the events in our catalogue.

\acknowledgments

We thank Pierre Fleury and Natalie B. Hogg for useful comments and discussions in the preparation of this work.
MM acknowledges funding by the Agenzia Spaziale Italiana (\textsc{asi}) under agreement no. 2018-23-HH.0 and support from INFN/Euclid Sezione di Roma. 
RM acknowledges financial contribution from Sapienza Universit\`a di Roma, thanks to Progetti di Ricerca Medi 2020, RM120172B32D5BE2 and 2021, RM12117A51D5269B 
and from the research grant number 2022E2J4RK ``PANTHEON: Perspectives in Astroparticle and Neutrino THEory with Old and New messengers''
under the program PRIN 2022 funded by the Italian Ministero dell’Universit\`a e della Ricerca (MUR). 

\bibliographystyle{JHEP}
\bibliography{references.bib}

\end{document}